\documentclass[11pt]{article}
\pdfoutput=1
\usepackage{jheppub}
\usepackage{blindtext}

\usepackage{mathrsfs}
\usepackage{bm}
\usepackage{paralist}
\usepackage{url}
\usepackage[]{microtype}

\usepackage{tikz-cd}
\usepackage{tikz}
\usepackage{pict2e}
\usepackage{float}
\usepackage{stmaryrd}
\usepackage{geometry}
\usepackage{tensor}
\usepackage{mathtools}

\usepackage{amsmath,amssymb,amsthm,amscd,graphicx}
\usepackage{psfrag}
\usepackage[utf8]{inputenc}
\usepackage{graphicx}
\usepackage{xcolor}
\usepackage{hyperref}
\hypersetup{colorlinks,allcolors=blue}
\definecolor{frangreen}{rgb}{0.040, 0.475, 0.435}

\usepackage[T1]{fontenc}

\usepackage[utf8]{inputenc}
\usepackage{graphicx}
\usepackage{amsmath,amsthm,amssymb,physics,mathtools}
\usepackage{tikz,tikz-cd, dynkin-diagrams}
\usepackage{epigraph}
\usepackage{hhline}
\usepackage{ytableau}
\usepackage{natbib}
\usetikzlibrary{positioning,arrows}
\usetikzlibrary{decorations.pathmorphing}
\usetikzlibrary{decorations.markings}
\usetikzlibrary{matrix}
\usepackage{bbold}

\usepackage{environ}         % provides \BODY
\usepackage{etoolbox}        % provides \ifdimcomp
\usepackage{graphicx}        % provides \resizebox
\usepackage{nomencl}
\makenomenclature

 \usepackage{etoolbox}
\renewcommand\nomgroup[1]{%
  \item[\bfseries
  \ifstrequal{#1}{C}{CFT symbols}{%
  \ifstrequal{#1}{H}{Heun symbols}{%
  \ifstrequal{#1}{F}{CFT symbols - semiclassics}{}}}%
]}
	\newlength{\myl}
\let\origequation=\equation
\let\origendequation=\endequation

\RenewEnviron{equation}{
  \settowidth{\myl}{$\BODY$}                       % calculate width and save as \myl
  \origequation
  \ifdimcomp{\the\linewidth}{>}{\the\myl}
  {\ensuremath{\BODY}}                             % True
  {\resizebox{\linewidth}{!}{\ensuremath{\BODY}}}  % False
  \origendequation
}

\theoremstyle{definition}

\newcommand{\be}{\begin{equation}}
\newcommand{\ee}{\end{equation}}
\newcommand{\ba}{\begin{aligned}}
\newcommand{\ea}{\end{aligned}}
\newcommand{\ben}{\begin{eqnarray}\displaystyle}
\newcommand{\een}{\end{eqnarray}}

\makeatletter
\gdef\@fpheader{}
\makeatother

\title{One loop corrections to the thermodynamics of near-extremal Kerr-(A)dS black holes from Heun equation}

\author[a]{Paolo Arnaudo}
\author[b]{Giulio Bonelli}
\author[c]{Alessandro Tanzini}

\affiliation[a]{Mathematical Sciences and STAG Research Centre, University of Southampton, Highfield, Southampton SO17 1BJ, UK}
\affiliation[b,c]{International School of Advanced Studies (SISSA), via Bonomea 265, 34136 Trieste, Italy}
\affiliation[b,c]{INFN, Sezione di Trieste}
\affiliation[b,c]{Institute for Geometry and Physics, IGAP, via Beirut 2, 34136  Trieste, Italy}

\emailAdd{P.Arnaudo@soton.ac.uk, bonelli@sissa.it, tanzini@sissa.it}
		
\date{\today}

\abstract{We compute one-loop corrections to the euclidean gravitational path integral of near-extremal (anti-)de Sitter-Kerr black hole in terms of the connection coefficients of the Heun equation describing the black hole linear perturbations in the Teukolsky formalism. 
We show that different near-extremal limits lead to distinct physical 
properties of the gravitational configuration, as they get described by distinct limiting differential equations. As a result, the light modes emerging in the limit determine different scaling properties in the temperature of the one-loop determinants. We show that the cold case displays distinctive universal log(T) corrections to the entropy of the system, including the ultracold regime. On the contrary, these do not appear in the  limit in which the event horizon superimposes onto the cosmological one.
In the Schwarzschild-de Sitter case, a further check is performed by comparison with the Denef-Hartnoll-Sachdev formula.
}

\begin{document}

\maketitle

\section{Introduction}

Recent developments in observational gravitational physics -- gravitational waves emitted by the merging of black holes (BH) binaries 
measured by the LIGO-Virgo-KAGRA collaboration and the BH images taken by the Event Horizon Telescope 
-- are opening a new observational window on gravity in the strong field and high dynamics regimes as well as to the near-horizon physics of supermassive black holes.
This further pushed BH physics as a more and more interesting subject to be explored at theoretical level.

Moreover, even if general relativity is understood as an effective theory, its quantum mechanical extension has been addressed via the Feynman path integral formalism in the Wilsonian sense \cite{PhysRevD.15.2752}. 
This is an open theoretical and mathematical problem due to several technical issues related to its implementation.

Although Einstein equations are given by PDEs, the mathematics of black hole perturbation theory (BHPT) gets simplified in ODEs
by separation of variables in presence of symmetries of the BH solutions of the gravitational equations \cite{teukolsky}.

To analyse the solutions of these ODEs, 
we use newly formulated techniques obtained by importing exact results from two-dimensional conformal field theory and supersymmetric gauge theory localization.
In this framework, the relevant solutions of the appropriate differential equations
are obtained from explicit combinatorial formul\ae \,
induced by the solution  of classical Virasoro conformal blocks. Once applied to BHPT,
this approach implies many results on BH physical observables, starting from
\cite{Aminov:2020yma,Bonelli:2021uvf,Bianchi:2021xpr}, see also \cite{CarneirodaCunha:2019tia, Bianchi:2021mft, Fioravanti:2021dce, Consoli:2022eey, Bautista:2023sdf, Cavalcante:2024kmy}, and in particular allows insight into the finite frequency perturbations.

In particular, these results can be used to compute one-loop corrections to the euclidean gravitational path integral, the effective determinant being computed in terms of the connection coefficients of the relevant ODEs. 
By making use of the relevant connection coefficients for the Heun equations as elaborated in \cite{Bonelli:2022ten},
this has been done in \cite{Arnaudo:2024rhv}, where we studied the on shell black hole one-loop effective actions for Kerr-de Sitter and Schwarzschild-(anti-)de Sitter black holes in four and five dimensions.

In the study of the gravitational path integral, recent conjectures on the appearance of light modes arising from Schwarzian reduction of 4d gravity was put forward in  \cite{Iliesiu:2020qvm,Iliesiu:2022onk,Kapec:2023ruw,Rakic:2023vhv,Banerjee:2023gll,Maulik:2024dwq,Kapec:2024zdj,Kolanowski:2024zrq,Maulik:2025phe}.
The above are based on the near-horizon approximation of near-extremal geometries,
the resulting spectrum displaying general common features both for asymptotically flat and (A)dS geometries \cite{Blacker:2025zca,Mariani:2025hee}. By using the previously mentioned techniques, in \cite{Arnaudo:2024bbd} we analysed the on-shell black hole perturbations
    directly in the full 4d metric for near-extremal Kerr BH in flat spacetime  
    and   
    found a set of 4d light modes that we expect to be related to the Schwarzian-like ones upon the near-horizon approximation.
    The scaling behavior in the temperature we found from the exact analysis of the full 4d spectrum contribution corroborates this expectation.
    In this paper, we continue this line of investigation by extending our analysis to near-extremal (A)dS$_4$ Kerr BHs. This is a much richer setup due to the existence of a further physical scale -- the cosmological constant -- and, in the de Sitter-Kerr case, different extremal geometries \cite{Romans:1991nq}. 
    
\begin{figure}[h!]
\centering
\includegraphics[width=16cm]{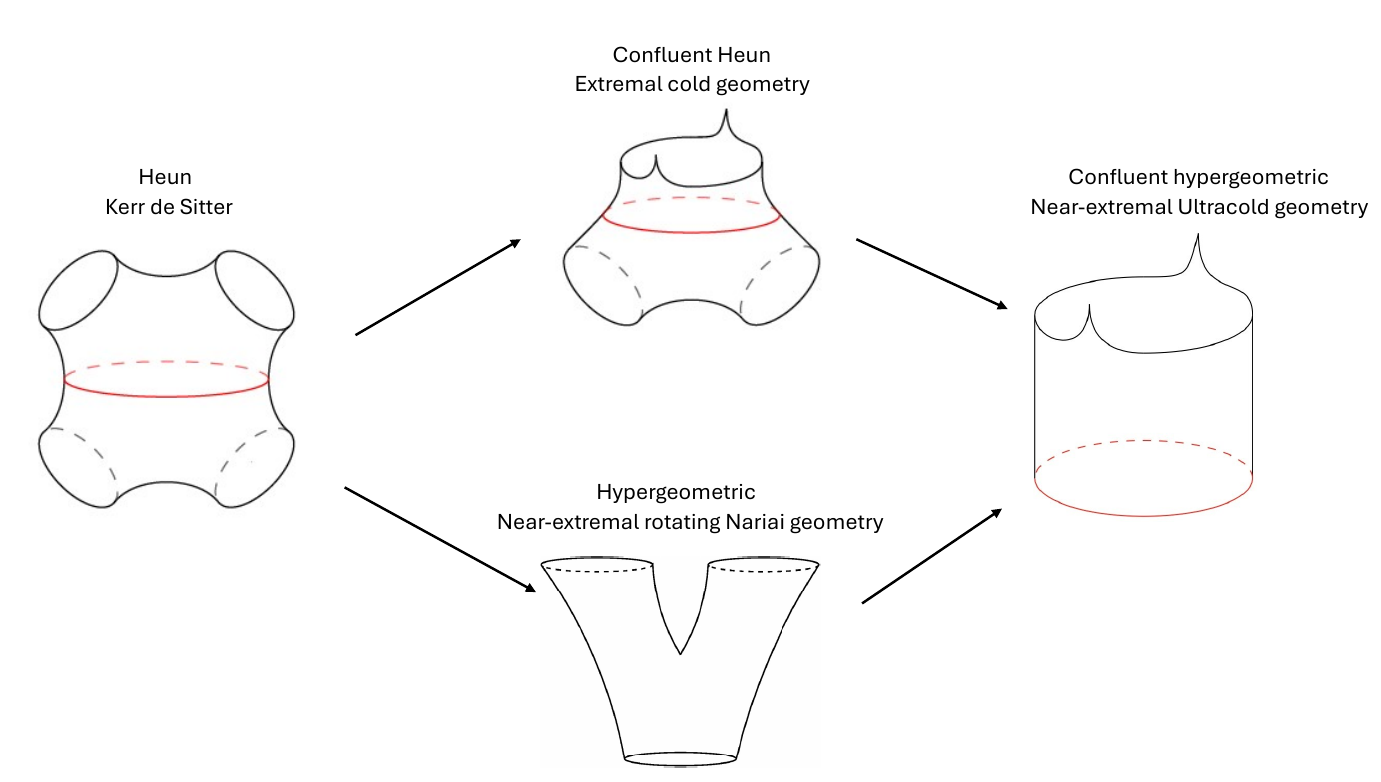}
\caption{Confluence diagram of ODEs for Kerr-dS radial problem and its (near-)extremal regimes. In the leftmost two pants surface, the four regular singularities appear. These correspond to the inner and outer horizon of the dS-Kerr BH, to the cosmological horizon and to a further (unphysical) zero of the metric profile function $\Delta_r(r)$. The upper left arrow indicates the cold extremal limit while the lower left one indicates the limit to rotating Nariai. The right arrows both lead to the ultracold BH configuration.}
\label{fig:confluences}
\end{figure}

     We analyse the problem by making use of the Teukolsky formalism to describe the physical degrees of freedom of the metric perturbations in terms of gauge invariant quantities. As already mentioned, in this context the perturbation equations separate into second-order ordinary differential equations both for the radial and the angular parts. For Kerr-(A)dS$_4$ these display four (regular) singularities and get identified to the regular Heun equation upon suitable dictionaries that we review in the main text.
    The singularity structure in the radial part corresponds to the geometry of the horizons of the BH under study.
     In particular, in the asymptotically de Sitter case, the different near-extremal limits are clearly distinct from the ODE viewpoint and the suitable boundary conditions, being represented either by confluencies or by superpositions of singularities.
    In Fig.\ref{fig:confluences} we represent the limits in this perspective, by denoting regular singularities with a 
    removed disk and an irregular singularity of rank one with a removed disk with two cusps, according to
    \cite{chekhov:hal-02501547}. See also \cite{Lisovyy:2021bkm, Lisovyy:2022flm} for more details and explanations.
    
\begin{figure}[h!]
\centering
\includegraphics[width=15cm]{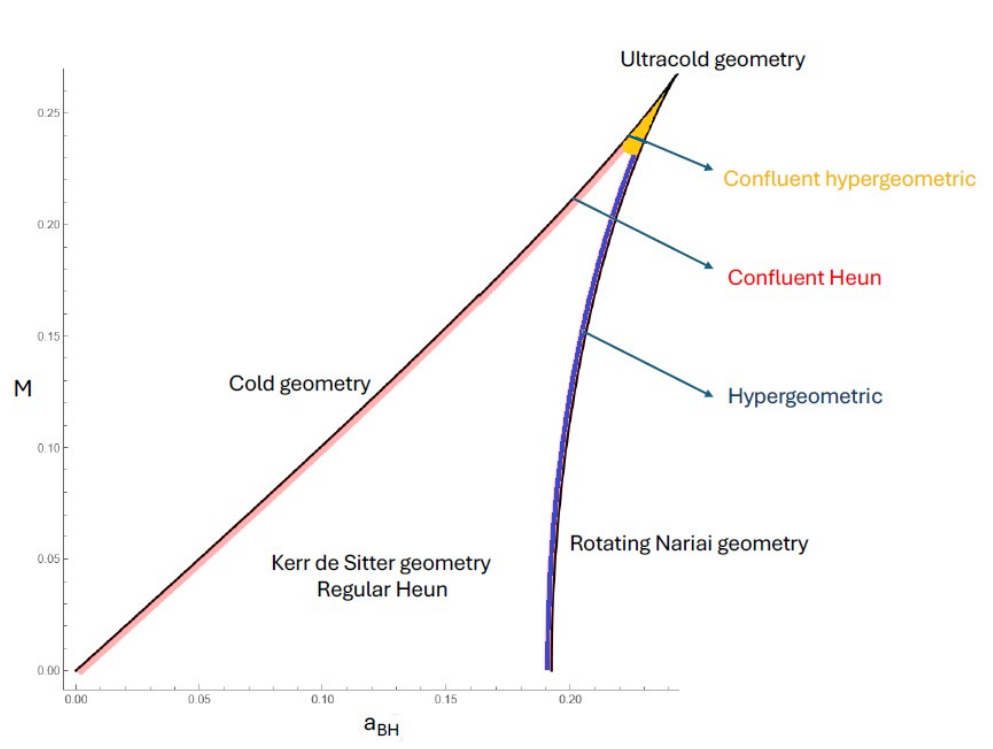}
\caption{Kerr-dS phase diagram: (Coloured) Romans shark fin picture \cite{Romans:1991nq}. Parameter space for Kerr-dS radial problem at $\Lambda=3$, with the various (near-)extremal geometries and the corresponding ODEs.}
\label{fig:sharkfin}
\end{figure}

        These different limits feature in turn the specific physical properties of the spectra of perturbations of the resulting geometries as dictated by the respective scaling of the gravitational parameters.
       The {\it confluence} limit is realised towards the cold BH geometry, while the {\it superposition} happens in the rotating Nariai one, see Fig.\ref{fig:sharkfin}.

 In particular,
    the contribution of the light modes can source logarithmic corrections to the BH entropy \cite{Banerjee:2010qc,Banerjee:2011jp,Sen:2012cj,Sen:2012kpz,  Bhattacharyya:2012ye,PandoZayas:2019hdb,Benini:2019dyp,Bobev:2023dwx}. 
    In the supersymmetric case, these are of the form $\log(A_H/G_N)$ where $A_H$ is the black hole horizon area. In the non-supersymmetric case the light modes originate in the near-horizon AdS$_2$ geometry and have been argued to be responsible for logarithmic corrections in the temperature to the entropy of near-extremal black holes
\cite{Iliesiu:2020qvm,Iliesiu:2022onk,Kapec:2023ruw,Rakic:2023vhv,Banerjee:2023gll,Maulik:2024dwq,Kapec:2024zdj,Kolanowski:2024zrq,Maulik:2025phe, Blacker:2025zca}.
    In the case of dS$_4$ BHs, the presence of log(T) corrections depend on the particular near-extremal limit under consideration. This is indeed very clear in our approach due to the distinct nature of the different limits of the ODE describing the black hole perturbation. More specifically, we find  
     that the presence of log(T) corrections is a distinctive feature of the {\it confluence} limit of the radial Teukolsky equation, describing cold near-extremal geometries. These corrections are analogous to the ones found in \cite{Arnaudo:2024bbd} in the asymptotically flat case. In particular, the result is universal, in the sense that depends only on the approaching inner and outer horizons, independently of the structure of the asymptotic geometry. 
     On the other hand, the {\it superposition} limit, describing the rotating Nariai geometry, does not produce any log(T) scaling of the one-loop effective action.

     The results of this analysis are further checked in the 
limiting non rotating Schwarzschild-de Sitter case, where the equations for the odd and even-parity gravitational perturbations can be analyzed. We also
compare with the DHS formula \cite{Denef:2009kn} which directly applies due to the static nature of the BH background geometry and the absence of branch cuts in the frequency plane. 
\footnote{
We notice that in the Reissner-N\"ordstrom case \cite{Blacker:2025zca} the limiting Nariai solution has negative specific heat. It may be interesting to investigate whether the same holds for the rotating Nariai geometry.
}
In Appendix \ref{appAdS} we also shortly discuss the universality of the cold extreme limiting case for AdS-Kerr, by studying it from the viewpoint of the QNMs quantization.
    
Let us list few further directions along which our results could be further developed.
In this paper we extended the analysis of \cite{Arnaudo:2024bbd} to Kerr black holes in (A)dS$_4$, proving the existence of a branch of light modes in the spectrum of gravitational perturbations at one-loop by a direct analysis of the latter in the full four-dimensional near-extremal background. This result is in line with approximated analysis in the near-horizon region of these or similar geometries \cite{Maulik:2025phe,Blacker:2025zca,Mariani:2025hee}.
An intriguing feature of the near-horizon analysis is the relation of the emergence of light modes to the reduction to a Schwarzian(-like) action for the metric. We expect the extension of these modes to the full four-dimensional geometry to be related to the light modes we find in our analysis. It would be very interesting to provide
a direct check of this conjecture. 
To this end, it would be useful also to extend our analysis to other black holes, such as Reissner-Nordstrom in (A)dS$_4$ 
and Myers-Perry \cite{10.21468/SciPostPhys.11.6.102,PhysRevD.108.124078}.
One technical point to elaborate is that while in the Schwarzian-like analysis the
linear perturbations of the metric are studied, our approach makes use of the Teukolsky formalism, where linear perturbations of gauge-invariant quantities build out from the curvature are considered. It is nonetheless possible to reconstruct metric perturbations from the latter by performing a suitable change of variables \cite{Shah:2012gu}. Moreover, it would be interesting to find a direct relation between the Liouville CFT at large central charge we use to solve the Heun equation and the Schwarzian theory, possibly generalising to Kerr the approach to rotating BTZ of \cite{Ghosh:2019rcj}.

It would also be important to extend our approach to supergravity in order to study black hole effective actions in this context and further check our formalism.

Other interesting aspects of the euclidean path integral approach to semiclassical quantum gravity that could be analyzed following our methods include the study of quantum cross-sections of near-extremal black holes \cite{Maulik:2025hax, Emparan:2025sao,Biggs:2025nzs} and the phase of the one-loop gravitational partition function \cite{Turiaci:2025xwi,Ivo:2025yek,Anninos:2025ltd,Law:2025yec}.
Related to this last point, we remark that in our analysis we do not take into account the superradiance instability of near-extremal Kerr black holes, and we consider only the absolute value of the one-loop effective action, as done in \cite{Rakic:2023vhv}.

\paragraph{\bf Acknowledgments:}
We would like to thank 
Nikolay Bobev,
Davide Cassani, 
Roberto Emparan, 
Sameer Murthy, 
Leopoldo Pando Zayas, 
Mukund Rangamani, 
Ashoke Sen,
Chiara Toldo 
and
Gustavo J. Turiaci 
for useful discussions, clarifications and/or a careful reading of the manuscript.
This research  is  partially supported by the INFN Research Projects GAST and ST$\&$FI, and by the PRIN project
"String theory as a bridge between Gauge Theories and Quantum Gravity".
All the authors acknowledge funding from the EU project Caligola (HORIZON-MSCA-2021-SE-01; Project ID: 101086123), and CA21109 - COST Action CaLISTA.
P.A. and A.T. also ackowledge GNFM group of INdAM.

\section{Kerr-de Sitter black hole in four dimensions}

The four-dimensional Kerr-de Sitter metric in Chambers-Moss coordinates reads
\begin{equation}\label{metricKerrdS}
\begin{aligned}
ds^2=\,&-\frac{\Delta_r(r)}{\left(r^2+a_{\text{BH}}^2\,\cos^2\theta\right)\left(1+\frac{a_{\text{BH}}^2\,\Lambda}{3}\right)^2}\left(\mathrm{d}t-a_{\text{BH}}\,\sin^2\theta\,\mathrm{d}\phi\right)^2\\
&+\frac{\left(1+\frac{a_{\text{BH}}^2\,\Lambda}{3}\cos^2\theta\right)\sin^2\theta}{\left(r^2+a_{\text{BH}}^2\,\cos^2\theta\right)\left(1+\frac{a_{\text{BH}}^2\,\Lambda}{3}\right)^2}\left[a_{\text{BH}}\,\mathrm{d}t-\left(r^2+a_{\text{BH}}^2\right)\mathrm{d}\phi\right]^2\\
&+\left(r^2+a_{\text{BH}}^2\,\cos^2\theta\right)\left(\frac{\mathrm{d}\theta^2}{1+\frac{a_{\text{BH}}^2\,\Lambda}{3}\cos^2\theta}+\frac{\mathrm{d}r^2}{\Delta_r(r)}\right),
\end{aligned}
\end{equation}
where
\begin{equation}\label{deltar}
\begin{aligned}
%u&=\cos\theta,\\
\Delta_r(r)&=r^2-2Mr+a_{\text{BH}}^2-\frac{\Lambda}{3}r^2(r^2+a_{\text{BH}}^2)=-\frac{\Lambda}{3}(r-R_+)(r-R_-)(r-R_h)(r-R_i).
\end{aligned}
\end{equation}
In \eqref{metricKerrdS}-\eqref{deltar}, $M$ is the mass of the black hole, $a_{\text{BH}}$ is its angular momentum, $\Lambda>0$ is the cosmological constant, and we have factorized the polynomial $\Delta_r(r)$ in linear terms, where $R_h$ is the event horizon, $R_i$ is the inner horizon, $R_{+}$ is the cosmological horizon, and $R_-<0$. 

In what follows, we will use both $R_j$-parameters -- as it is useful to visualize the confluences happening in the realizations of the (near-)extremal geometries -- and the parameters $a_{\text{BH}}, M, \Lambda$. Of course, these parameters are not independent. For example, an easy and useful relation is
\begin{equation}
R_-=-R_+-R_i-R_h.
\end{equation}
Moreover, we can express the parameters $M,\Lambda,a_{\text{BH}}$ in terms of $R_h,R_i,R_+$ as follows:
\begin{equation}
\begin{aligned}
M&= -\frac{(R_h+R_i) (R_h+R_+) (R_i+R_+)}{4 R_h R_i R_+ (R_h+R_i+R_+)}\\
&\quad\times\biggl[-\sqrt{\left(R_h^2+R_h (R_i+R_+)+R_i^2+R_i R_++R_+^2\right)^2+4 R_h R_i R_+ (R_h+R_i+R_+)}+R_h^2+R_h (R_i+R_+)+R_i^2+R_i R_++R_+^2\biggr],\\
\Lambda &= \frac{6}{\sqrt{\left(R_h^2+R_h (R_i+R_+)+R_i^2+R_i R_++R_+^2\right)^2+4 R_h R_i R_+ (R_h+R_i+R_+)}+R_h^2+R_h (R_i+R_+)+R_i^2+R_i R_++R_+^2},\\
a_{\text{BH}}^2&= \frac{\sqrt{\left(R_h^2+R_h (R_i+R_+)+R_i^2+R_i R_++R_+^2\right)^2+4 R_h R_i R_+ (R_h+R_i+R_+)}-R_h^2-R_h (R_i+R_+)-R_i^2-R_i R_+-R_+^2}{2}.
\end{aligned}
\end{equation}
It will be useful to have also the relation giving $R_{j_2}\in\{R_i,R_+\}$ and $a_{\text{BH}}$ in terms of $R_{j_1}\in \{R_i,R_+\}\setminus \{R_{j_2}\}$, $R_h$, and $\Lambda$:
\begin{equation}
\begin{aligned}
R_{j_2}&= -\frac{R_h+R_{j_1}}{2}+\frac{1}{2} \sqrt{\frac{12}{\Lambda }-\left(4 a_{\text{BH}}^2+3 R_h^2+2 R_h R_{j_1}+3 R_{j_1}^2\right)} ,\\
a_{\text{BH}}^2&=\frac{R_h R_{j_1} \left(3-\Lambda  R_h^2-\Lambda  R_h R_{j_1}-\Lambda  R_{j_1}^2\right)}{\Lambda  R_h R_{j_1}+3}.
\end{aligned}
\end{equation}

%We used units in which $c=G=1$, the coordinates $t$ and $r$ range over $]-\infty,\infty[$, and $\theta,\phi$ are standard coordinates on $S^2$.

Teukolsky \cite{teukolsky} showed that using the so-called \emph{Newman-Penrose formalism}, it is possible to derive the separable master equations for the scalar, electromagnetic, and gravitational perturbations of the Kerr black hole. An analogous separation of variables can be applied to Kerr-(Anti)-de Sitter black holes \cite{khanal}. 
Introducing the Teukolsky wave function $\Phi_s$, defined from the Newman-Penrose scalar quantities, and imposing the Ansatz
\begin{equation}
\Phi_s(t,r,u,\phi)=e^{i(-\omega t+m\phi)}R_s(r)S_s(u),
\end{equation}
where $u=\cos\theta$, the master equation separates in the following radial and angular equations:
\begin{equation}
\label{radialKerr}
\begin{aligned}
&\Delta_r^{-s}(r)\frac{d}{dr}\biggl(\Delta_r^{s+1}(r)\frac{dR(r)}{dr}\biggr)+\Biggl[\frac{[\omega(r^2+a_{\text{BH}}^2)-a_{\text{BH}}\,m]^2\bigl(1+\frac{\Lambda}{3}a_{\text{BH}}^2\bigr)^2-is\Delta'_r(r)[\omega(r^2+a_{\text{BH}}^2)-a_{\text{BH}}\,m]\bigl(1+\frac{\Lambda}{3}a_{\text{BH}}^2\bigr)}{\Delta_r(r)}\\
&\ \ \ +4is\omega\left(1+\frac{\Lambda}{3}a_{\text{BH}}^2\right)r-\frac{2\Lambda}{3}(s+1)(2s+1)r^2+s\left(1-\frac{\Lambda}{3}a_{\text{BH}}^2\right)-{}_sA_{\ell m}\Biggr]R(r)=0,
\end{aligned}
\end{equation}
\begin{equation}
\begin{aligned}
&\frac{d}{du}\biggl[\left(1-u^2\right) \left(1+\frac{a_{\text{BH}}^2 \Lambda}{3}\,  u^2\right)\frac{dS(u)}{du}\biggr]+\Biggl[-\frac{\bigl[\bigl(1+\frac{a_{\text{BH}}^2\Lambda}{3}\bigr)[m-a_{\text{BH}}\omega(1-u^2)]+su\bigl(\frac{a_{\text{BH}}^2\Lambda}{3}  \left(2 u^2-1\right)+1\bigr)\bigr]^2}{(1-u^2)\biggl(1+\frac{a_{\text{BH}}^2 \Lambda}{3}u^2\biggr)}\\
&\ \ \ -4s\,a_{\text{BH}}\,\omega\left(1+\frac{a_{\text{BH}}^2\Lambda}{3}\right)u-\frac{2a_{\text{BH}}^2\Lambda}{3}(2s^2+1)u^2+{}_sA_{\ell m}\Biggr]S(u)=0,
\end{aligned}
\end{equation}
where with ${}_sA_{\ell m}$ we denote the separation constant. 

As noticed in \cite{Suzuki:1998vy}, both equations can be written as Heun differential equations. Moreover, in the limit in which the cosmological constant vanishes, these reproduce the radial and angular equations of the same class of perturbations in asymptotically flat Kerr spacetime, and both become confluent Heun equations.
The confluence procedure is analogous to the one analyzed in Appendix \ref{appheun}.

\section{Angular problem}

The singularities of the angular equation are located at the following points in the $u$-plane:
\begin{equation}
\pm 1,\quad\pm \frac{i}{a_{\text{BH}}}\, \sqrt{\frac{3}{\Lambda }}.
\end{equation}
If we define the new variable
\begin{equation}
z=\frac{\frac{2i}{a_{\text{BH}}}\, \sqrt{\frac{3}{\Lambda }}}{1+\frac{i}{a_{\text{BH}}}\, \sqrt{\frac{3}{\Lambda }}}\cdot\frac{u+1}{u+\frac{i}{a_{\text{BH}}}\, \sqrt{\frac{3}{\Lambda }}},
\end{equation}
and we redefine the wave function as
\begin{equation}
S(u)=\left(z-z_{\infty}\right)z^{-\theta_0}(z-1)^{-\theta_1}(z-t)^{-\theta_t}w(z),
\end{equation}
where
\begin{equation}
\begin{aligned}
t&=\frac{\frac{4i}{a_{\text{BH}}}\, \sqrt{\frac{3}{\Lambda }}}{\left(1+\frac{i}{a_{\text{BH}}}\, \sqrt{\frac{3}{\Lambda }}\right)^2},\quad\quad
z_{\infty}=\frac{\frac{2i}{a_{\text{BH}}}\, \sqrt{\frac{3}{\Lambda }}}{1+\frac{i}{a_{\text{BH}}}\, \sqrt{\frac{3}{\Lambda }}},\\
\theta_0&=\frac{m-s}{2},\quad\quad\quad \theta_t=-\frac{m+s}{2},\\
\theta_1&=-\frac{s}{2}+\frac{i}{2} \sqrt{\frac{3}{\Lambda }} \left[\frac{a_{\text{BH}}\,\Lambda}{3}  \,(a_{\text{BH}}\, \omega -m)+\omega \right],
\end{aligned}
\end{equation}
the differential equation becomes a Heun equation
\eqref{heuncanonical}
with
\begin{equation}
\begin{aligned}
%t&=\frac{4 a_{\text{BH}} \sqrt{-\frac{3}{\Lambda }}}{\left(a_{\text{BH}}+i\,\sqrt{\frac{3}{\Lambda }}\right)^2},\\
\alpha&=1+s+i\,\sqrt{\frac{3}{\Lambda }} \left[\frac{a_{\text{BH}} \Lambda}{3}\,  (m-a_{\text{BH}} \omega )-\omega \right],\\
\beta&=1+2s,\\
\gamma&=1+s-m,\\
\delta&=1+s-i\,\sqrt{\frac{3}{\Lambda }} \left[\frac{a_{\text{BH}} \Lambda}{3}\,  (a_{\text{BH}} \omega -m)+\omega \right],\\
\epsilon&=1+s+m;\\
q&=\frac{\Lambda  ({}_sA_{\ell m}-s)-2 a_{\text{BH}}^3 \Lambda  (2 s+1) \omega +a_{\text{BH}}^2 [2 \Lambda  m+s (4 \Lambda  m+3)]+2 a_{\text{BH}} (2 s+1) \left[\Lambda  \omega +i\,\sqrt{3 \Lambda } (1+s-m)\right]}{\left(a_{\text{BH}}+i\,\sqrt{\frac{\Lambda }{3}}\right)^2}.
\end{aligned}
\end{equation}

For the angular problem we impose as boundary conditions the regularity of the solutions at $\theta=0,\pi$, which correspond to $u=\pm 1$, and so to $z=0$ and $z=t$.
We remark that $t$ is close to 0 if the dimensionless parameter $a_{\text{BH}}\,\sqrt{\Lambda}$ is small. In this case, the solution for the separation constant can be found by imposing the following quantization to the gauge parameter $a$, parametrizing the composite monodromy around the singularities $z=0$ and $z=t$:
\begin{equation}
a=\ell+\frac{1}{2},\ \ \text{where}\ \ \ell\ge\max\{|s|,|m|\}\ \ \text{and}\ \ell\in\mathbb{N},
\end{equation}
since we are considering integer spin perturbations \footnote{For half-integer spin perturbations, $\ell\ge\max\{|s|,|m|\}$ and $\ell\in\mathbb{N}+\frac{1}{2}$.}.

The expansion of the separation constant is usually obtained as a double expansion in $a_{\text{BH}}^2\,\Lambda$ and $a_{\text{BH}}\,\omega$ \cite{Novaes:2018fry, Suzuki:1998vy}. At leading order in $a_{\text{BH}}^2\,\Lambda$ and first order in $a_{\text{BH}}\,\omega$, we find 
\begin{equation}
\begin{aligned}
{}_sA_{\ell m}&=\ell(\ell+1)-s^2-\frac{2m[\ell(\ell+1)+s^2]}{\ell(\ell+1)}\,a_{\text{BH}}\,\omega+\mathcal{O}(a_{\text{BH}}\,\sqrt{\Lambda}),
\end{aligned}
\end{equation}
which agrees with \cite{Novaes:2018fry}.

We remark that by taking $\omega=0$, the equation remains a Heun equation. Therefore, the angular problem does not simplify in the (near-)extremal regimes, which only affect the singularity structure of the radial equation. The only modification appears in the expression of the separation constant. In what follows, we will always denote the separation constant as ${}_sA_{\ell m}$, although different regimes for $\omega$ (that will be specified case-by-case) have to be understood. 
Instead, by taking $\Lambda\to 0$ with $a_{\text{BH}}$ fixed, the equation reduces to a confluent Heun equation, encoding the angular problem for the four-dimensional flat Kerr black hole.
Finally, by taking $a_{\text{BH}}\to 0$ with $\Lambda$ fixed, the equation reduces to a Legendre/hypergeometric equation, and the separation constant simply reduces to $\ell(\ell+1)-s^2$. 

The procedure in which the Heun equation gives rise to confluent Heun or to hypergeometric equations is described in Appendix \ref{appheun}.

\section{Radial problem}

In this section, we consider the differential equation \eqref{radialKerr} and, in particular, we analyze its near-extremal regimes.
It is useful to introduce the temperature and angular velocity at the four horizons:
\begin{equation}
\begin{aligned}
T_j&=\frac{|\Delta'_r(R_j)|}{4 \pi  \left(1+\frac{\Lambda}{3}a_{\text{BH}}^2\right) \left(a_{\text{BH}}^2+R_j^2\right)},\\
\Omega_j&=\frac{a_{\text{BH}}}{a_{\text{BH}}^2+R_j^2},
\end{aligned}
\end{equation}
for $j\in\{i,h,+,-\}$.

The extremal geometries can be visualized as curves limiting the region of physically allowed parameters in the space of all parameters. We present the space of admissible $a_{\text{BH}}, M$ parameters at fixed $\Lambda=3$ in Figure \ref{fig:sharkfin}. %Another way to visualize the picture is to introduce the dimensionless parameters $\alpha\equiv a_{\text{BH}}/M$ and $\beta\equiv a_{\text{BH}}^2 \Lambda$  (see Figure \ref{fig:sharkfin2}).

%\begin{figure}[h!]
%    \centering
%    \includegraphics[width=80mm]{sharkfin1.pdf}
%    \caption{Space of admissible $a_{\text{BH}}, M$ parameters at fixed $\Lambda=3$ for Kerr-dS$_4$ geometry. On the horizontal axis $a_{\text{BH}}$, on the vertical axis $M$.}
%    \label{fig:sharkfin1}
%\end{figure}
%\begin{figure}[h!]
%    \centering
%    \includegraphics[width=80mm]{sharkfin2.pdf}
%    \caption{On the horizontal axis $a_{\text{BH}}/M$, on the vertical axis $a_{\text{BH}}^2\Lambda$.}
%    \label{fig:sharkfin2}
%\end{figure}

We first consider two cases of near-extremal geometries, the case in which $R_i\sim R_h$, which we refer to as the \emph{near-cold geometry}, analysed in Section \ref{sec:coldKerrdS}, and the one in which $R_+\sim R_h$, which we refer to as the \emph{rotating Nariai geometry}. The cold geometry is represented by the line through the origin in Figure \ref{fig:sharkfin}, whereas the rotating Nariai geometry is represented by the region close to the curve on the right of the picture. The intersection point of the line with the curve represents the ultracold geometry, which can be obtained by setting $R_i=R_h=R_+$. We analyse the near-ultracold regime starting from the rotating Nariai one in Section \ref{sec:ultracold}. As we will see in the following discussion, the first two cases have different properties, both from a physical and a mathematical viewpoint.

\subsection{Near-cold geometry}\label{sec:coldKerrdS}

In the near-cold geometry,
we introduce the variable
\begin{equation}\label{zKerrdScold}
z=\frac{(r-R_i) (R_+-R_-)}{(r-R_-) (R_+-R_i)},
\end{equation}
and we redefine the wave function as
\begin{equation}
R(z)=\frac{(-R_i z+R_-+R_+ (z-1))^{2 s+1}}{ \left\{z(1-z)\left[z (R_h-R_-) (R_i-R_+)-(R_h-R_i) (R_--R_+)\right]\right\}^{\frac{s+1}{2} }}\psi(z).
\end{equation}
The resulting differential equation is a Heun equation in normal form \eqref{heundiff},
with dictionary
\begin{equation}\label{dictioKerrdScold} 
\begin{aligned}
t&=\frac{(R_h-R_i) (R_+-R_-)}{(R_h-R_-) (R_+-R_i)},\\
a_0&=\frac{s}{2}-\frac{i (\omega -m \Omega_i)}{4 \pi\,T_i},\\
a_t&=\frac{s}{2}+\frac{i (\omega -m \Omega_h)}{4 \pi\,T_h},\\
a_1&=\frac{s}{2}-\frac{i (\omega -m \Omega_+)}{4 \pi\,T_+},\\
a_{\infty}&=\frac{s}{2}+\frac{i (\omega -m \Omega_-)}{4 \pi\,T_-},
\end{aligned}
\end{equation}
\begin{equation}
\begin{aligned}
u&=-\frac{s \left(a_{\text{BH}}^2 \Lambda -3\right)}{\Lambda  (R_h-R_+) (R_h+2 R_i+R_+)}+\frac{3 {}_sA_{\ell m}}{\Lambda  (R_h-R_+) (R_i-R_-)}\\
&-\frac{1}{2 \Lambda ^2 (R_h-R_i)^2 (R_h-R_-) (R_h-R_+)^3 (R_i-R_-)}\biggl\{36 \left(a_{\text{BH}} m-\omega  \left(a_{\text{BH}}^2+R_h^2\right)\right)\\
&\times\left[-a_{\text{BH}}^2 \omega  (3 R_h+R_-)+a_{\text{BH}} m (3 R_h+R_-)-R_h \omega  (R_h (R_i+R_+)-2 R_i R_+)\right]\\
&+6 \Lambda  \biggl[4 a_{\text{BH}}^6 \omega ^2 (3 R_h+R_-)-8 a_{\text{BH}}^5 m \omega  (3 R_h+R_-)+4 a_{\text{BH}}^4 \left(m^2 (3 R_h+R_-)+2 R_h \omega ^2 \left(R_h^2-R_i R_+\right)\right)\\
&-8 a_{\text{BH}}^3 m R_h \omega  \left(R_h^2-R_i R_+\right)+2 a_{\text{BH}}^2 \omega  \left(2 R_h^3 \omega  (R_h (R_i+R_+)-2 R_i R_+)-i s (R_h-R_i)^2 (R_h-R_+)^2\right)\\
&+2 i a_{\text{BH}} m s (R_h-R_i)^2 (R_h-R_+)^2+2 s (R_h-R_i)^2 (R_h-R_+)^2 (R_h (2-i R_- \omega )+R_i+R_+)\biggr]\\
&+\Lambda ^2 \biggl[4 a_{\text{BH}}^8 \omega ^2 (3 R_h+R_-)-8 a_{\text{BH}}^7 m \omega  (3 R_h+R_-)+4 a_{\text{BH}}^6 \left(m^2 (3 R_h+R_-)+2 R_h \omega ^2 \left(R_h^2-R_i R_+\right)\right)\\
&-8 a_{\text{BH}}^5 m R_h \omega  \left(R_h^2-R_i R_+\right)-4 a_{\text{BH}}^4 \omega  \left(R_h^3 \omega  (2 R_i R_+-R_h (R_i+R_+))+i s (R_h-R_i)^2 (R_h-R_+)^2\right)\\
&+4 i a_{\text{BH}}^3 m s (R_h-R_i)^2 (R_h-R_+)^2-4 a_{\text{BH}}^2 s (R_h-R_i)^2 (R_h-R_+)^2 (R_h (2+i R_- \omega )+R_i+R_+)\\
&+(s+1) (R_h-R_i)^2 (R_h-R_-) (R_h-R_+)^2 \left(-s \left(-2 R_h R_-+R_i^2+R_+^2\right)-(R_i+R_+)^2\right)\biggr]\biggr\}.
\end{aligned}
\end{equation}
The relevant connection formula is the one between $z=t$ and $z=1$ in the regime $t\sim 0$.
The boundary conditions select the local solutions
\begin{equation}
\begin{aligned}
\psi(z)&=\psi_{t,-}(z)\sim (z-t)^{\frac{1}{2}-a_t},\quad z\sim t,\\
\psi(z)&=\psi_{1,+}(z)\sim (z-1)^{\frac{1}{2}+a_1},\quad z\sim 1.
\end{aligned}
\end{equation}
From the physical viewpoint, these select the ingoing solution at the outer horizon $R_h$ and the outgoing solution at the cosmological horizon $R_+$.
The connection formula analytically continuing the solution $\psi_{t,-}(z)$ in the region $z\sim 0$ reads
\begin{equation}
\begin{aligned}
\psi_{t,-}(z)=\sum_{\theta=\pm}\left(\sum_{\sigma=\pm}\frac{\Gamma\left(1-2a_t\right)\Gamma\left(-2\sigma a\right)\Gamma\left(1-2\sigma a\right)\Gamma\left(-2\theta a_1\right)}{\prod_{\pm}\Gamma\left(\frac{1}{2}-a_t-\sigma a\pm a_0\right)\Gamma\left(\frac{1}{2}-\sigma a-\theta a_1\pm a_{\infty}\right)}t^{\sigma a }e^{-\frac{\sigma}{2}\partial_aF}\right)t^{-a_t}e^{-\frac{1}{2}\partial_{a_t}F-\frac{\theta}{2}\partial_{a_1}F}\psi_{1,\theta}(z),
\end{aligned}
\end{equation}
where $F\equiv F^{N_f=4}(t)$ denotes the Nekrasov-Shatashvili (NS) free energy of the $\mathcal{N}=2$ $SU(2)$ gauge theory with 4 (anti-)fundamental hypermultiplets. We refer to Appendix \ref{appgauge} for the definition of the relevant quantities and the conventions used.

Therefore, the quantization condition for the QNM frequencies can be found by setting to zero the connection coefficient in front of the discarded solution:
\begin{equation}
\sum_{\sigma=\pm}\frac{\Gamma\left(-2\sigma a\right)\Gamma\left(1-2\sigma a\right)\Gamma\left(2 a_1\right)}{\prod_{\pm}\Gamma\left(\frac{1}{2}-a_t-\sigma a\pm a_0\right)\Gamma\left(\frac{1}{2}-\sigma a+ a_1\pm a_{\infty}\right)}t^{\sigma a }e^{-\frac{\sigma}{2}\partial_aF}=0.
\end{equation}
Among the possible solutions, there are long-lived modes that can be found from the poles of the $\Gamma$-function whose argument diverges in the regime $R_i\sim R_h$, that is
\begin{equation}\label{divergingGammas}
\frac{1}{2}+a+a_0-a_t=-n,\quad n\in\mathbb{Z}_{\ge 0},
\end{equation}
which gives
\begin{equation}\label{lowlyingQNMs}
\omega_n=m\,\Omega_h+m\,\frac{a_{\text{BH}}\,R_h}{\left(a_{\text{BH}}^2+R_h^2\right)^2}(R_h-R_i)-i\,\pi (2n+1+2 a^{(0)}(m\,\Omega_h))T_h+\mathcal{O}(T_h^2),
\end{equation}
where we used the following notation for the small temperature expansion of the composite monodromy parameter\footnote{We remark that the instanton expansion parameter $t$ is proportional to $R_h-R_i$, so that $T_h$ and $t$ scale in the same way in the near-extremal regime.} 
\begin{equation}
a\equiv a(\omega)=a^{(0)}(\omega)+a^{(1)}(\omega)\,T_h+\mathcal{O}(T_h^2),
\end{equation}
and in \eqref{lowlyingQNMs} we evaluated $\omega$ at its leading order $m\,\Omega_h$.

In the strict $T_h\to 0$ limit, realized by $R_i\to R_h$, the resulting differential equation is a confluent Heun equation, and the $\Gamma$-function with divergent argument decouples from the resulting connection problem \cite{Bonelli:2022ten}. The confluence procedure is described in Appendix \ref{appheun}.

Similarly to the asymptotically flat Kerr geometry \cite{Arnaudo:2024bbd}, the $\log T_h$-behavior can be found considering the $\Gamma$-function $\Gamma(\frac{1}{2}-a_t+a_0-a)$ and evaluating it in the Matsubara frequencies.
The latters can be found by imposing 
\begin{equation}
2a_t=-k,\quad k\in\mathbb{Z}_{\ge 0},
\end{equation}
which gives
\begin{equation}\label{Matsubara}
\omega_k^{(M)}=m\,\Omega_h+2\pi i (k+s)T_h.
\end{equation}
Evaluating $\frac{1}{2}-a+a_0-a_t$ in \eqref{Matsubara} and expanding for $R_i\sim R_h$, we find that the contributions to the temperature only arise when $m=0$: indeed, as in the flat case \cite{Arnaudo:2024bbd}, we need $m=0$ to ensure that the leading order of the argument of the $\Gamma$-function can be a nonpositive integer.
Substituting 
\begin{equation}
\begin{aligned}
R_-&=-R_+-R_h-R_i,\\
R_{+}&= -\frac{R_h+R_i}{2}+\frac{1}{2} \sqrt{\frac{12}{\Lambda }-\left(4 a_{\text{BH}}^2+3 R_h^2+2 R_h R_i+3 R_i^2\right)} ,\\
a_{\text{BH}}^2&=\frac{R_h R_i \left(-\Lambda  R_h^2-\Lambda  R_h R_i-\Lambda  R_i^2+3\right)}{\Lambda  R_h R_i+3},
\end{aligned}
\end{equation}
we can expand $\frac{1}{2}-a+a_0-a_t$ around $\Lambda=0$ \footnote{The same result would also hold using an expansion around $a_{\text{BH}}\sim 0$, that is, in the small rotation regime, which coincides with the perturbation around the pure de Sitter case, as can be deduced from Figure \ref{fig:sharkfin}.} and we find that the argument of the $\Gamma$-function reduces to
\begin{equation}\label{argumentexpansion1}
\frac{1}{2}+k+s-\frac{1}{2}\sqrt{1+4s^2+4{}_sA_{\ell m}^{(0)}}-\frac{k+s}{2 R_h}(R_h-R_i)+\mathcal{O}\left[(R_h-R_i)^2\right],
\end{equation}
where ${}_sA_{\ell m}^{(0)}$ is the leading order of the expansion of the separation constant.
Using that in the small $\Lambda$ and $\omega$ expansions\footnote{The expansion is around small $\omega$, because we are evaluating the argument of the $\Gamma$-function in the Matsubara frequencies and at $m=0$, therefore their contribution is proportional to the temperature $T_h$.} 
\begin{equation}
{}_sA_{\ell m}^{(0)}=\ell(\ell+1)-s^2,
\end{equation}
the expansion \eqref{argumentexpansion1} reduces to
\begin{equation}
k+s-\ell-2\pi\,(k+s)\,(R_h+R_i)\,T_h+\mathcal{O}\left(T_h^2\right).
\end{equation}
The procedure then follows the same steps described in \cite{Arnaudo:2024bbd}. In particular, the temperature scaling for perturbations of spin $s=1,2$ can be found from
\begin{equation}\label{infiniteproducts}
\begin{aligned}
&\prod_{\ell\ge s}\ \prod_{k\ge 0}\Gamma\left(k+s-\ell-2\pi\,(k+s)\,(R_h+R_i)\,T_h\right)=\prod_{\ell'\ge 0}\ \prod_{k\ge 0}\Gamma\left(k-\ell'-2\pi\,(k+s)\,(R_h+R_i)\,T_h\right)=\\
&\prod_{k\ge 0}\ \prod_{r\in\mathbb{Z}}\Gamma\left(r-2\pi\,(k+s)\,(R_h+R_i)\,T_h\right)\sim \prod_{k\ge 0}\ \prod_{r\in\mathbb{Z}}\ \prod_{n\ge 0}\left(n+r-2\pi\,(k+s)\,(R_h+R_i)\,T_h\right)^{-1}=\\
&\prod_{r\in\mathbb{Z}}\prod_{ n+r\ne 0}\prod_{k\ge 0}\left(n+r-2\pi\,(k+s)\,(R_h+R_i)\,T_h\right)^{-1}\prod_{n\ge 0}\prod_{k\ge 0}\left(-2\pi\,(k+s)\,(R_h+R_i)\,T_h\right)^{-1},
\end{aligned}
\end{equation}
and, selecting the last product,
\begin{equation}\label{secondstepinfiniteproducts}
\begin{aligned}
&\prod_{n\ge 0}\prod_{k\ge 0}\left(-2\pi\,(k+s)\,(R_h+R_i)\,T_h\right)^{-1}\sim\\
&\prod_{n\ge 0}\Gamma_1\left(2\pi\,s\,(R_h+R_i)\,T_h\mid 2\pi\,(R_h+R_i)\,T_h\right)=\\
&\prod_{n\ge 0}\frac{\Gamma(s)}{\sqrt{2\pi}}\left(2\pi\,(R_h+R_i)\,T_h\right)^{s-\frac{1}{2}}\sim
T_h^{\frac{s}{2}-\frac{1}{4}}.
\end{aligned}
\end{equation}
Taking into account the contributions from the QNM and anti-QNM frequencies, and both signs of $s$ for $|s|= 1,2$, we find that the gravitational one-loop partition function scales as $T_h^{3/2}$ and the electromagnetic one as $T_h^{1/2}$.

\subsection{Rotating Nariai geometry}

In the rotating Nariai geometry, 
the same procedure analysed in the near-cold case holds, and both the transformations and the dictionary can be obtained by switching $R_+\leftrightarrow R_i$. The main difference between the two cases lies in the form of the quantization condition for the QNMs. 

Indeed, in this case the relevant connection formula is the one between $z=0$ and $z=t$ in the regime $t\sim 0$.
The boundary conditions select the local solutions
\begin{equation}
\begin{aligned}
\psi(z)&=\psi_{t,-}(z)\sim (z-t)^{\frac{1}{2}-a_t},\quad z\sim t,\\
\psi(z)&=\psi_{0,+}(z)\sim z^{\frac{1}{2}+a_0},\quad z\sim 0.
\end{aligned}
\end{equation}

The relevant connection formula reads
\begin{equation}\label{connectionNariai}
\begin{aligned}
\psi_{0,+}(z)=\sum_{\theta=\pm}\frac{\Gamma\left(1+2a_0\right)\Gamma\left(-2\theta a_t\right)}{\prod_{\pm}\Gamma\left(\frac{1}{2}+a_0-\theta a_t\pm a\right)}t^{a_0-\theta a_t}e^{\frac{1}{2}\partial_{a_0}F-\frac{\theta}{2}\partial_{a_t}F}\psi_{t,\theta}(z),
\end{aligned}
\end{equation}
from which the QNM frequencies are found by imposing
\begin{equation}
\frac{\Gamma\left(1+2a_0\right)\Gamma\left(-2 a_t\right)}{\prod_{\pm}\Gamma\left(\frac{1}{2}+a_0-a_t\pm a\right)}t^{a_0-a_t}e^{\frac{1}{2}\partial_{a_0}F-\frac{1}{2}\partial_{a_t}F}=0.
\end{equation}
The analysis for the Matsubara frequencies is exactly the same as seen in the previous case, and the same result \eqref{Matsubara} holds.

Similarly to the near-cold-regime, we evaluate $\frac{1}{2}-a+a_0-a_t$ in the Matsubara frequencies, and expand for $R_+\sim R_h$. The difference, however, is that in this case we cannot hit any pole of the $\Gamma$-function. For example,
in the small $a_{\text{BH}}$ expansion\footnote{We expect the small $a_{\text{BH}}$ assumption to be non-restrictive for the analysis, since the reduction to the near-extremal regime never involves the mechanism of decoupling of a $\Gamma$-function when approaching the near-extremal regime. In addition to this, it is the regime that allows us to make contact with the Schwarzschild-de Sitter case analysed in Section \ref{sec:SdS} which is simpler since it depends on less parameters.}, the leading order reads  
\begin{equation}
\frac{1}{2}+k+s-\frac{1}{2} \sqrt{\frac{1}{3} \left(8 s^2-5\right)-4 \ell (\ell+1)},
\end{equation}
which, for $s=2$, is 
\begin{equation}
\frac{5}{2}+k-\frac{1}{2} \sqrt{9-4 \ell (\ell+1)}.
\end{equation}
In particular, $\sqrt{9-4 \ell (\ell+1)}$ is imaginary since $\ell\ge 2$. Therefore, the leading order never hits a pole of the $\Gamma$-function and we do not find contributions proportional to the temperature.

Let us motivate our result from the point of view of the differential equation. In this regime, the strict $R_+\to R_h$ limit is not meaningful for the perturbation problem, since it would force the (incompatible) boundary conditions to be imposed at the same point. The near-extremal regime can be reached by appropriately redefining the radial variable and reducing the differential equation to a hypergeometric one. Let us define
\begin{equation}
z=\frac{(R_+-R_h)(r-R_-)}{(R_+-R_-)(r-R_h)}.
\end{equation}
The radial differential equation admits a well-defined leading order in the small $(R_+-R_h)$-expansion if the frequency is redefined as
\begin{equation}
\omega=\frac{m\,a_{\text{BH}}}{(a_{\text{BH}}^2+R_h^2)^2}+\hat{\omega}(R_+-R_h).
\end{equation}
In this way, redefining also the wave function as
\begin{equation}
R(z)=z^s\,(1-z)^{-\frac{s+1}{2}}\,\psi(z),
\end{equation}
the resulting differential equation at leading order in $(R_+-R_h)$ reads
\begin{equation}\label{ODEnearrotatingNariai}
\psi''(z)+V_h(z)\,\psi(z)=0,
\end{equation}
with
\begin{equation}
\begin{aligned}
V_h(z)=&-\frac{2 \left(a_{\text{BH}}^2-1\right) s+{}_sA_{\ell m}+2 R_h^2 (s+1) (2 s+1)}{z^2 (z-1) (R_h-R_i) (3 R_h+R_i)}+\frac{4 a_{\text{BH}}^2 \left(a_{\text{BH}}^2+1\right)^2 m^2 R_h^2}{(z-1)^2 z^2 \left(a_{\text{BH}}^2+R_h^2\right)^2 (R_h-R_i)^2 (3 R_h+R_i)^2}\\
&+\frac{2 i a_{\text{BH}} \left(a_{\text{BH}}^2+1\right) m R_h s}{(z-1)^2 z \left(a_{\text{BH}}^2+R_h^2\right) (R_h-R_i) (3 R_h+R_i)}+\frac{4 a_{\text{BH}} \left(a_{\text{BH}}^2+1\right)^2 m R_h \hat{\omega}}{(z-1)^2 z (R_h-R_i)^2 (3 R_h+R_i)^2}\\
&-\frac{i \left(a_{\text{BH}}^2+1\right) s \hat{\omega} (z-2) \left(a_{\text{BH}}^2+R_h^2\right)}{(z-1)^2 z (R_h-R_i) (3 R_h+R_i)}+\frac{\left(a_{\text{BH}}^2+1\right)^2 \hat{\omega}^2 \left(a_{\text{BH}}^2+R_h^2\right)^2}{(z-1)^2 (R_h-R_i)^2 (3 R_h+R_i)^2}-\frac{(s+1) \left(s (z-2)^2-z^2\right)}{4 (z-1)^2 z^2},
\end{aligned}
\end{equation}
which is of hypergeometric type \eqref{Hypergeometricnormal}, which indicial parameters
\begin{equation}
\begin{aligned}
a_0&=\frac{1}{2 \left(a_{\text{BH}}^2+R_h^2\right) \left(-3 R_h^2+2 R_h R_i+R_i^2\right)}\biggl[8 a_{\text{BH}}^6 \left(-2 m^2 R_h^2-s (R_h-R_i) (3 R_h+R_i)\right)+a_{\text{BH}}^4 \biggl(-32 m^2 R_h^2\\
&-4 {}_sA_{\ell m} (R_h-R_i) (3 R_h+R_i)-(R_h-R_i) (3 R_h+R_i) \left(R_h^2 (4 s (s+7)+5)+2 R_h R_i (2 s+1)^2+(2 R_i s+R_i)^2-8 s\right)\biggr)\\
&+2 a_{\text{BH}}^2 R_h^2 \left(-(R_h-R_i) (3 R_h+R_i) \left(4 {}_sA_{\ell m}+R_h^2 (4 s (s+4)+5)+2 R_h R_i (2 s+1)^2+(2 R_i s+R_i)^2-8 s\right)-8 m^2\right)\\
&-R_h^4 (R_h-R_i) (3 R_h+R_i) \left(4 {}_sA_{\ell m}+(2 s+1) \left(5 R_h^2+2 s (R_h+R_i)^2+2 R_h R_i+R_i^2\right)-8 s\right)\biggr]^{1/2},\\
a_1&=\frac{s}{2}-\frac{i \left(a_{\text{BH}}^2+1\right) \left(\hat{\omega} \left(a_{\text{BH}}^2+R_h^2\right)^2+2 a_{\text{BH}} m R_h\right)}{\left(a_{\text{BH}}^2+R_h^2\right) (R_h-R_i) (3 R_h+R_i)},\\
a_{\infty}&=\frac{-R_h^2 \left(3 s+2 i \left(a_{\text{BH}}^2+1\right) \hat{\omega}\right)-2 i \left(a^2+1\right) a_{\text{BH}}^2 \hat{\omega}+2 R_h R_i s+R_i^2 s}{2 \left(-3 R_h^2+2 R_h R_i+R_i^2\right)}.
\end{aligned}
\end{equation}
In particular, passing from the full problem described by the Heun equation to the near-extremal regime described by the hypergeometric equation, the reduction of the connection problem does not involve the decoupling of a $\Gamma$-function, as happened in the cold-limit. We describe in Appendix \ref{appheun} the superposition/ungauging procedure through which the Heun equation reduces to a hypergeometric one.

As a further confirmation, in the next section we consider the near-extremal regime of the Schwarzschild-de Sitter black hole.
The analysis of the non-rotating case is simpler thanks to the fewer parameters involved and to the applicability of the DHS formula \cite{Denef:2009kn}, see also \cite{Mukherjee:2025xlt}.

\section{Schwarzschild-de Sitter in four dimensions}\label{sec:SdS}

In this section, we consider the gravitational perturbations of the Schwarzschild-de Sitter black hole in four dimensions, in the near-extremal regime, to have a simpler testing ground for the result of the previous section.

The metric of the Schwarzschild-de Sitter black hole in four dimensions is given by
\begin{equation}
\mathrm{d}s^2=-f(r)\,\mathrm{d}t^2+\frac{\mathrm{d}r^2}{f(r)}+r^2\,\mathrm{d}\Omega_2^2,
\end{equation}
with
\begin{equation}
f(r)=1-\frac{2M}{r}-\frac{\Lambda}{3}r^2=-\frac{\Lambda}{3}\frac{(r-R_h)(r-R_+)(r-R_-)}{r},
\end{equation}
where 
\begin{equation}
R_{\pm}=\frac{-R_h\pm \sqrt{\frac{12}{\Lambda }-3 R_h^2}}{2}.
\end{equation}
Similarly to the Kerr-de Sitter case, $R_+$ denotes the cosmological horizon, while $R_-$ is a negative real root.

The gravitational perturbations around Schwarzschild-de Sitter black holes can be divided in odd and even-parity perturbations (see for example \cite{Cardoso:2001bb, Kodama:2003jz}), and are encoded in differential equations that in the flat limit $\Lambda\to 0$ reduce to the well-known Regge-Wheeler \cite{RW} and Zerilli \cite{zerilli} equations. The analysis for both cases is similar, but the singularity structure of the equation for even-parity perturbations is more complicated, and we postpone it to Appendix \ref{appzerilli}.

\subsection{Odd-parity gravitational perturbations}\label{sec:RW}

The differential equation describing odd-parity gravitational perturbations is given by
\begin{equation}\label{RWSdS}
R''(r)+\frac{f'(r)}{f(r)}R'(r)+\frac{\omega ^2-V(r)}{f(r)^2}R(r)=0,
\end{equation}
where the potential is 
\begin{equation}
V(r)=f(r) \left(\frac{\ell (\ell+1)}{r^2}-\frac{6 M}{r^3}\right).
\end{equation}
We first address the problem of obtaining the Matsubara frequencies and the QNM frequencies  by using the Heun connection formulae.
Defining the new variable
\begin{equation}
z=\frac{R_-}{R_h}\frac{(r-R_h) }{(r-R_-)},
\end{equation}
and redefining the wave function as
\begin{equation}
R(z)=\frac{\sqrt{1-z}}{\sqrt{z} \sqrt{R_h R_- (z-1)-R_h R_+ z+R_- R_+}}\,\psi(z),
\end{equation}
we find that $\psi(z)$ satisfies the Heun equation in normal form \eqref{heundiff} with dictionary
\begin{equation}
\begin{aligned}
t&=\frac{R_-}{R_h}\frac{(R_+-R_h) }{(R_+-R_-)},\\
a_0&=\frac{3 i R_h \omega }{\Lambda  (R_h-R_-) (R_h-R_+)},\\
a_t&=\frac{3 i R_+ \omega }{\Lambda  (R_h-R_+) (R_--R_+)},\\
a_1&=2,\\
a_{\infty}&=\frac{3 i R_- \omega }{\Lambda  (R_h-R_-) (R_+-R_-)},\\
u&=\frac{\Lambda  (6 \ell (\ell+1)+\Lambda  R_h (7 R_--R_+))+\frac{36 R_h R_+^2 \omega ^2}{(R_h-R_+)^2 (R_--R_+)}}{2 \Lambda ^2 R_+ (R_h-R_-)}.
\end{aligned}
\end{equation}
The local solutions selected by the boundary conditions are $\psi_+^{(0)}(z)$ and $\psi_-^{(t)}(z)$, so that the relevant connection problem is the same seen in the rotating Nariai geometry \eqref{connectionNariai}.

In the near-extremal limit, we see that, %the gauge mass going to infinity 
as $R_h\sim R_+$,  
\begin{equation}
a_0-a_t=-\frac{2 i \omega }{\Lambda  (R_+-R_h)}+\mathcal{O}((R_+-R_h)^0),
\end{equation}
leading to the divergence of the argument of the $\Gamma$-functions involved in the connection coefficients.

The QNM frequencies are found by imposing 
\begin{equation}
\frac{1}{2}+a+a_0-a_t=-n,
\end{equation}
with the rescaling
\begin{equation}\label{omegahatSdS}
\hat{\omega}\equiv \frac{\omega}{(R_+-R_h)},
\end{equation}
needed to obtained well defined results. At leading order in $(R_+-R_h)$, (and using also that in the near-extremal regime $\Lambda\to 1/R_h^2$) we find 
\begin{equation}\label{QNMSdSCardoso}
2 R_h^2\hat{\omega}_n=\frac{\omega_n }{\frac{R_+-R_h}{2 R_h^2}}= \frac{1}{2} \sqrt{4 \ell (\ell+1)-9}-\frac{i}{2}\, (2 n+1),
\end{equation}
in accordance with \cite{Cardoso:2003sw}.

The Matsubara frequencies are given by the condition
\begin{equation}
-2a_0=-k,\quad k\in\mathbb{Z}_{\ge 0},
\end{equation}
from which
\begin{equation}\label{MatsubaraSdS}
\omega_k^{(M)}=2i\pi\, k\,  T_h=i\, k\,\frac{f'(R_h)}{2}=\frac{i k \Lambda  (R_h-R_-) (R_+-R_h)}{6 R_h},
\end{equation}
where we used that 
\begin{equation}\label{temperatureSdS}
T_h=f'(R_h)/(4\pi).
\end{equation}

We remark that in the near-extremal regime, the temperature \eqref{temperatureSdS} reduces to
\begin{equation}
T=(R_+-R_h)/(4\pi\,R_h^2),
\end{equation}
so that \eqref{QNMSdSCardoso} can be rewritten as
\begin{equation}\label{QNMSdS2}
\omega_n=2\,\pi\,T_h\left(\frac{1}{2}\sqrt{4\ell(\ell+1)-9}-\frac{i}{2}(2n+1)\right),
\end{equation}
and \eqref{MatsubaraSdS} reduces to
\begin{equation}
\omega_k^{(M)}=\frac{i\,k}{2R_h^2}(R_+-R_h).
\end{equation}

To address the temperature scaling of the one-loop partition function, we evaluate the divergent argument $\frac{1}{2}-a+a_0-a_t$ in the Matsubara frequencies, and, using $R_-=-R_+-R_h$, we find
\begin{equation}\label{nearNariaiSdS}
\frac{1}{2}-a+a_0-a_t=\frac{1}{2}+k-\frac{1}{2} \sqrt{9-\frac{4 \ell (\ell+1)}{\Lambda  R_h^2}}+\mathcal{O}((R_+-R_h)^1).
\end{equation}
Using also that $\Lambda\to 1/R_h^2$ near-extremality, \eqref{nearNariaiSdS} reduces to
\begin{equation}\label{nopole}
\frac{1}{2}+k-\frac{1}{2} \sqrt{9-4 \ell (\ell+1)},
\end{equation}
that, since $\ell\ge 2$, always has an imaginary part, never hitting a pole of the $\Gamma$-function. This is in accordance with the result found in the rotating Nariai case in the previous section.

By using the explicit expression for the frequencies \eqref{QNMSdS2}, we can give an additional (but equivalent, as argued in \cite{Arnaudo:2024rhv}) argument for the missing scaling with the temperature. Indeed, since the Schwarzschild-de Sitter black hole is a static spacetime and no branch cuts are present in the frequency plane, the DHS formula applies, and we can use Equation (41) in \cite{Denef:2009kn} to see the contribution of the QNMs to the partition function:
\begin{equation}
\begin{aligned}
&\prod_{k\ge 0}\left(k+\frac{i \omega_n}{2\pi T}\right)=\prod_{k\ge 0}\left(k+\frac{2i \pi T\left(\frac{1}{2}\sqrt{4\ell(\ell+1)-9}-\frac{i}{2}(2n+1)\right)}{2\pi T}\right)=\\
&\prod_{k\ge 0}\left(k+ \left(\frac{1}{2}\sqrt{9-4\ell(\ell+1)}+\frac{1}{2}(2n+1)\right)\right),
\end{aligned}
\end{equation}
where we inserted \eqref{QNMSdS2} for the QNM frequencies.
We see that the temperature dependence goes away since these leading orders are always nonzero (for every $k,n\ge 0$): indeed, $\sqrt{9-4\ell(\ell+1)}$ is an imaginary number, since $\ell\ge 2$. The same reasoning applies to the anti-QNMs. This is in accordance with the observation that \eqref{nopole} never hits a pole of the $\Gamma$-function.

We finally remark that by defining a different variable $z$, it is possible to rewrite the differential equation at small $(R_+-R_h)$ in hypergeometric form, as seen in the rotating Nariai case. This permits us once more to see explicitly the ungauging procedure we describe in Appendix \ref{appheun}.  We define a variable $z$ in such a way that the singularities which overlap are at $z=t$ and $z=\infty$. 
Defining 
\begin{equation}
z=\frac{(r-R_h) (R_+-R_-)}{(r-R_-) (R_+-R_h)},
\end{equation}
and redefining the wave function in \eqref{RWSdS} as
\begin{equation}
R(z)=\frac{\sqrt{R_h R_- z-R_h R_-+R_h R_+-R_- R_+ z}}{\sqrt{1-z} \sqrt{z}}\,\psi(z),
\end{equation}
we find that $\psi(z)$ satisfies the Heun equation in normal form \eqref{heundiff} with dictionary
\begin{equation}
\begin{aligned}
t&=\frac{R_h (R_+-R_-)}{R_- (R_+-R_h)},\\
a_0&=\frac{3 i R_h \omega }{\Lambda  (R_h-R_-) (R_h-R_+)},\\
a_1&=\frac{3 i R_+ \omega }{\Lambda  (R_h-R_+) (R_--R_+)},\\
a_t&=2,\\
a_{\infty}&=\frac{3 i R_- \omega }{\Lambda  (R_h-R_-) (R_+-R_-)},\\
u&=\frac{\Lambda  \left(7 R_h^2+22 R_h R_++7 R_+^2\right)-6 \ell (\ell+1)}{2 \Lambda  R_+ (2 R_h+R_+)}.
\end{aligned}
\end{equation}
We note that, in the limit $R_+\to R_h$, we have
\begin{equation}
u\to 6-\frac{\ell (\ell+1)}{\Lambda \, R_h^2},\quad a_{\infty}\to \frac{2 i \omega }{3 \Lambda  R_h},
\end{equation}
whereas $t,\,a_0,\,a_1\to\infty$.
If we want the leading order in the small $(R_+-R_h)$-expansion to be smooth, we have to redefine again the frequency as in \eqref{omegahatSdS}.
In this way, the Heun differential equation simplifies at leading order to
\begin{equation}
\psi''(z)+\frac{\Lambda  \left(4 (z-1) z \left(\ell(\ell+1)-2 \Lambda  R_h^2\right)+\Lambda  R_h^2\right)+4 R_h^2 \omega ^2}{4 \Lambda ^2 R_h^2 (z-1)^2 z^2}\,\psi(z)=0,
\end{equation}
which is a hypergeometric differential equation with indicial parameters
\begin{equation}
a_0=a_1=\frac{i \hat{\omega} }{\Lambda },\quad a_{\infty}=\frac{\sqrt{9 \Lambda  R_h^2-4 \ell (\ell+1)}}{2 \sqrt{\Lambda } R_h}.
\end{equation}
Using also that, near-extremality, $\Lambda\to 1/R_h^2$, we see that $a_{\infty}\to \frac{1}{2} \sqrt{9-4 \ell (\ell+1)}$, and this is precisely the leading order of the v.e.v. $a$ as in the above computation. Indeed, after the superposition of the singularities at $z=t$ and $z=\infty$, the composite monodromy parameter $a$ simplifies to the indicial parameter $a_{\infty}$.

\section{Near-Ultracold geometry}\label{sec:ultracold}

In the Kerr-de Sitter geometry, there is one additional near-extremal geometry which is worth considering, that is the near-ultracold solution. In the extremal regime, the geometry develops a two-dimensional Minkowski near-horizon geometry, and corresponds to the limit in which $R_h=R_+=R_i$.

To study the near-ultracold regime, we start from the rotating Nariai geometry, and take the additional approximation $R_h\sim R_i$. With this approach, we are investigating the region close to the point in which the two extremal curves intersect in the parameter space plotted in Figure \ref{fig:sharkfin}. In particular, the limit $R_i\to R_h$ corresponds to moving towards the extremal cold limit by starting from the rotating Nariai one. Therefore, we expect the mechanism of decoupling of a $\Gamma$-function in the connection formulae to take place, and hence the differential equation to be reduced to a confluent hypergeometric one. 

Let us start by considering the hypergeometric differential equation \eqref{ODEnearrotatingNariai} describing the perturbation equation in the rotating Nariai geometry.
The relevant connection problem is the one between $z=1$, which corresponds to the cosmological horizon, and $z=\infty$, which corresponds to the event horizon. In particular, the selected local solutions, are
\begin{equation}
\begin{aligned}
\psi_+^{(1)}(z)\sim (z-1)^{\frac{1}{2}+a_1}\left[1+\mathcal{O}(z-1)\right],\\
\psi_-^{(\infty)}(z)\sim z^{\frac{1}{2}-a_\infty}\left[1+\mathcal{O}(1/z)\right].
\end{aligned}
\end{equation}
The quantization condition for the quasinormal modes amounts to finding poles in the $\Gamma$-functions in the hypergeometric connection formulae, and in this case are found by imposing\footnote{This is the reduction of the connection formula (4.1.23) in \cite{Bonelli:2022ten} in the limit $t\to 0$ which gives the well known connection formula between the hypergeometric functions centered at $z=1$ and the one at $z=\infty$, following the superposition procedure explained in Appendix \ref{appheun}.}
\begin{equation}
\frac{1}{2}+a_1+a_{\infty}\pm a_0=-n,\quad n\in\mathbb{Z}_{\ge 0}.
\end{equation}
Among the two combinations, we have that $\frac{1}{2}+a_1+a_{\infty}+a_0$ is finite in the $R_i\to R_h$ limit, whereas $\frac{1}{2}+a_1+a_{\infty}-a_0$ is divergent.  Moreover, the potential of the differential equation \eqref{ODEnearrotatingNariai} is divergent in the $R_i\to R_h$ limit, and it is necessary to further rescale the frequency as 
\begin{equation}
\hat{\omega}=(R_h-R_i)\,\hat{\hat{\omega}}
\end{equation}
to cure the divergence.
Indeed, defining the new variable
\begin{equation}
y=\frac{i \left(a_{\text{BH}}+a_{\text{BH}}^3\right) m}{z \left(a_{\text{BH}}^2+R_h^2\right) (R_h-R_i)},
\end{equation}
and redefining the wave function as
\begin{equation}
\psi(y)=e^{-\frac{y}{2}} y^{\frac{i \,\hat{\hat{\omega}}\left(a_{\text{BH}}^2+1\right)  \left(a_{\text{BH}}^2+R_h^2\right)}{4R_h}+ \frac{s-1}{2}}\,w(y),
\end{equation}
the differential equation at leading order in $(R_h-R_i)$ becomes
\begin{equation}
\begin{aligned}
&y\, w''(y)+\left[\frac{i \left(a_{\text{BH}}^2+1\right) a_{\text{BH}}^2 \hat{\hat{\omega}}}{2 R_h}+\frac{1}{2} i \left(a_{\text{BH}}^2+1\right) R_h \hat{\hat{\omega}}+s-y+1\right]w'(y)\\
&-\frac{i}{4 a_{\text{BH}} \left(a_{\text{BH}}^2+1\right) m R_h} \biggl[{}_sA_{\ell m} \left(-a_{\text{BH}}^2-R_h^2\right)+2 a_{\text{BH}} \left(a_{\text{BH}}^2+1\right)^2 m \hat{\hat{\omega}} \left(a^2+R_h^2\right)\\
&-2 i a_{\text{BH}} \left(a_{\text{BH}}^2+1\right) m R_h-2 \left(a_{\text{BH}}^2+R_h^2\right) \left(\left(a_{\text{BH}}^2-1\right) s+R_h^2 (s+1) (2 s+1)\right)\biggr]w(y)=0,
\end{aligned}
\end{equation}
which is indeed a confluent hypergeometric equation.
We notice that in the above computations, we rescaled twice the frequency and the radial variable, since the temperature $T_h$ is proportional to the product $(R_h-R_i)(R_+-R_h)$. This is in accordance with the discussion in \cite{Mariani:2025hee}, after Equation (3.18), where the authors point out that in the ultracold regime the temperature is of order two in the decoupling parameter, differently from the cold and the rotating Nariai cases, in which the temperature scales linearly. In our case, we see explicitly the dependence on the product of the two small quantities.
Furthermore, the scaling with the temperature follows the same mechanism seen in Section \ref{sec:coldKerrdS}: indeed, the divergent argument $\frac{1}{2}+a_1+a_{\infty}-a_0$ of the $\Gamma$-function is the analog of the argument in \eqref{divergingGammas}, in the new (rescaled) variable and in the hypergeometric regime, in which the composite monodromy $a$ is replaced by the indicial parameter of one of the singularities (here $a_0$).

We finally remark that, thanks to the symmetry in $R_+$ and $R_i$ of the original differential equation, the same procedure can be repeated by first considering the small $(R_h-R_i)$ regime and then the small $(R_+-R_h)$ one.

\appendix

\section{Even-parity gravitational perturbations in four-dimensional Schwarzschild-de Sitter }\label{appzerilli}

In this Appendix, we repeat the analysis done in Section \ref{sec:RW} for the even-parity gravitational perturbations.

Following the notations in Section 3 of \cite{Kodama:2003jz},
the differential equation is given by
\begin{equation}
\begin{aligned}
&R''(r)+\frac{2 \left(\Lambda  r^3-3 M\right)}{r \left(6 M+\Lambda  r^3-3 r\right)}\,R'(r)+\biggl[\frac{9 r^2 \omega ^2}{\left(6 M+\Lambda  r^3-3 r\right)^2}\\
&+\frac{3 \left(6 \left(\ell^2+\ell-2\right)^2 M r^2+\ell (\ell+1) \left(\ell^2+\ell-2\right)^2 r^3-12 M^2 r \left(-3 \ell (\ell+1)+2 \Lambda  r^2+6\right)+72 M^3\right)}{r^2 \left(\left(\ell^2+\ell-2\right) r+6 M\right)^2 \left(6 M+\Lambda  r^3-3 r\right)}\biggr]R(r)=0.
\end{aligned}
\end{equation}
This has five regular singularities: the three roots of $r\,f(r)=0$ given by $R_h,R_{\pm}$ as in the odd-parity case, $r=0$, and the unphysical point $r=\frac{6 M}{2-\ell (\ell+1)}$.
Defining the new variable
\begin{equation}
z=\frac{6 M(r-R_h)}{r (2-\ell (\ell+1)) \left(\frac{6 M}{2-\ell (\ell+1)}-R_h\right)},
\end{equation}
and redefining the wave function as
\begin{equation}
R(z)=\frac{1}{\sqrt{z} \sqrt{\left(k^2-2\right) R_h R_- z+6 M (R_h+R_- (z-1))} \sqrt{\left(k^2-2\right) R_h R_+ z+6 M (R_h+R_+ (z-1))}}\,\psi(z),
\end{equation}
we find that $\psi(z)$ satisfies the 5-point Fuchsian differential equation in normal form 
\begin{equation}
\psi''(z)+\left[\frac{-\frac{3}{4}-a_{\infty}^2+a_0^2+a_1^2-+a_q^2+a_t^2}{z (z-1)}+\frac{\frac{1}{4}-a_0^2}{z^2}+\frac{\frac{1}{4}-a_1^2}{(z-1)^2}+\frac{\frac{1}{4}-a_q^2}{(z-q)^2}+\frac{\frac{1}{4}-a_t^2}{(z-t)^2}+\frac{(q-1) u_q}{z (z-1) (z-q)}+\frac{(t-1) u_t}{z (z-1) (z-t)}\right]\psi(z)=0,
\end{equation}
with dictionary 
\begin{equation}\nonumber
\begin{aligned}
t&=\frac{6 M (R_+-R_h)}{R_+ \left(2-\ell(\ell+1)\right) \left(\frac{6 M}{2-\ell(\ell+1)}-R_h\right)},\quad\quad
q=\frac{6 M (R_--R_h)}{R_- \left(2-\ell(\ell+1)\right) \left(\frac{6 M}{2-\ell(\ell+1)}-R_h\right)},\\
a_0&=\frac{3 i R_h \omega }{\Lambda  (R_h-R_-) (R_h-R_+)},\quad\quad\quad\quad
a_t=\frac{3 i R_+ \omega }{\Lambda  (R_h-R_+) (R_--R_+)},\quad\quad
a_{\infty}=0,
\end{aligned}
\end{equation}
\begin{equation}\nonumber
\begin{aligned}
a_1&=\sqrt{\frac{\Lambda  \left(\ell^2+\ell-2\right)^2 \left(R_h^2+R_h R_++R_+^2\right)+3 \left(\ell^2+\ell-2\right)^2 (3 \ell (\ell+1)+2)-9 \Lambda ^3 R_h^2 R_+^2 (R_h+R_+)^2}{4 \left(\ell^2+\ell-\Lambda  R_h R_+-2\right) \left(\ell^2+\ell+\Lambda  R_h (R_h+R_+)-2\right) \left(\ell^2+\ell+\Lambda  R_+ (R_h+R_+)-2\right)}},
\end{aligned}
\end{equation}
\begin{equation}
\begin{aligned}
u_t&=\frac{1}{2 \Lambda ^2 R_h (R_h-R_+)^2 (R_h+2 R_+)^3 \left(\ell^2+\ell+\Lambda  R_h (R_h+R_+)-2\right)^2}\biggl[6 \Lambda  \ell^3 (\ell+1)^3 (R_h-R_+)^2 (R_h+2 R_+)^2\\
&+36 R_+^2 \omega ^2 \left(-2 R_h^2-2 R_h R_++R_+^2\right) \left(\ell^2+\ell+\Lambda  R_h (R_h+R_+)-2\right)^2+\Lambda  \ell^2 (\ell+1)^2 (R_h-R_+)^2 (R_h+2 R_+)^2 \left(\Lambda  (2 R_h+R_+)^2-24\right)\\
&+2 \Lambda  \ell (\ell+1) (R_h-R_+)^2 (R_h+2 R_+)^2 (\Lambda  R_h (R_h+R_+)-2) \left(\Lambda  \left(R_h^2+R_h R_++R_+^2\right)-6\right)\\
&-4 \Lambda ^3 R_h^8-3 \Lambda ^4 R_h^8 R_+^2-12 \Lambda ^4 R_h^7 R_+^3-16 \Lambda ^3 R_h^7 R_++16 \Lambda ^2 R_h^6-6 \Lambda ^4 R_h^6 R_+^4-12 \Lambda ^3 R_h^6 R_+^2\\
&+24 \Lambda ^4 R_h^5 R_+^5+20 \Lambda ^3 R_h^5 R_+^3+48 \Lambda ^2 R_h^5 R_++21 \Lambda ^4 R_h^4 R_+^6+32 \Lambda ^3 R_h^4 R_+^4-12 \Lambda ^2 R_h^4 R_+^2\\
&-12 \Lambda ^4 R_h^3 R_+^7+12 \Lambda ^3 R_h^3 R_+^5-104 \Lambda ^2 R_h^3 R_+^3-12 \Lambda ^4 R_h^2 R_+^8-16 \Lambda ^3 R_h^2 R_+^6-12 \Lambda ^2 R_h^2 R_+^4-16 \Lambda ^3 R_h R_+^7+48 \Lambda ^2 R_h R_+^5+16 \Lambda ^2 R_+^6\biggr],\\
u_q&=-\frac{1}{2 \Lambda ^2 R_h (2 R_h+R_+)^2 (R_h+2 R_+)^3 \left(\ell^2+\ell-\Lambda  R_h R_+-2\right)^2}\biggl[6 \Lambda  \ell^3 (\ell+1)^3 (2 R_h+R_+)^2 (R_h+2 R_+)^2\\
&+36 \omega ^2 (R_h+R_+)^2 \left(R_h^2+4 R_h R_++R_+^2\right) \left(\ell^2+\ell-\Lambda  R_h R_+-2\right)^2\\
&+\Lambda  \ell^2 (\ell+1)^2 (2 R_h+R_+)^2 (R_h+2 R_+)^2 \left(\Lambda  (R_h-R_+)^2-24\right)\\
&-2 \Lambda  \ell (\ell+1) (2 R_h+R_+)^2 (R_h+2 R_+)^2 (\Lambda  R_h R_++2) \left(\Lambda  \left(R_h^2+R_h R_++R_+^2\right)-6\right)\\
&-12 \Lambda ^4 R_h^8 R_+^2-84 \Lambda ^4 R_h^7 R_+^3+16 \Lambda ^3 R_h^7 R_++16 \Lambda ^2 R_h^6-231 \Lambda ^4 R_h^6 R_+^4+96 \Lambda ^3 R_h^6 R_+^2-318 \Lambda ^4 R_h^5 R_+^5\\
&+228 \Lambda ^3 R_h^5 R_+^3+48 \Lambda ^2 R_h^5 R_+-231 \Lambda ^4 R_h^4 R_+^6+292 \Lambda ^3 R_h^4 R_+^4-12 \Lambda ^2 R_h^4 R_+^2-84 \Lambda ^4 R_h^3 R_+^7\\
&+228 \Lambda ^3 R_h^3 R_+^5-104 \Lambda ^2 R_h^3 R_+^3-12 \Lambda ^4 R_h^2 R_+^8+96 \Lambda ^3 R_h^2 R_+^6-12 \Lambda ^2 R_h^2 R_+^4+16 \Lambda ^3 R_h R_+^7+48 \Lambda ^2 R_h R_+^5+16 \Lambda ^2 R_+^6\biggr].
\end{aligned}
\end{equation}
The gauge theory encoding the above spectral problem is the $\mathcal{N}=2$ $SU(2)\times S(2)$ quiver gauge theory analyzed in \cite{Arnaudo:2025kof}.

We remark that the expressions of $a_0$ and $a_t$ are the same as in the odd-parity case, so that also the relevant connection problem and the selected local solutions are the same. Moreover, at leading order in the instanton expansions, the composite monodromy parameter around $z=0$ and $z=t$ is given by
\begin{equation}
b_1=\sqrt{-\frac{1}{4}+a_0^2+a_t^2- u_t}+\mathcal{O}(t).
\end{equation}
%\PA{Here, $b_1$ is the analogous of $a$, and is the intermediate region adjacent to the patch containing $z=0$ and $z=t$. I chose the plus sign as always, and the correction at 1st instanton is a correction in $t$ which is proportional to $(R_+-R_h)$.}

The Matsubara frequencies also admits the same expressions \eqref{MatsubaraSdS}, as in the odd-parity case, since they only depend on $a_0$.
By substituting the Matsubara frequencies \eqref{MatsubaraSdS} in the argument of the $\Gamma$-function
\begin{equation}
\Gamma\left(\frac{1}{2}-b_1+a_0-a_t\right),
\end{equation}
%expanding in $R_+-R_h$, and using again that $R_-=-R_+-R_h$ and that in the extremal case $\Lambda=1/R_h^2$, 
we find that at leading order near-extremality
\begin{equation}
\frac{1}{2}-b_1+a_0-a_t=\frac{1}{2}+k-\frac{1}{2} \sqrt{9-4 \ell (\ell+1)}+\mathcal{O}((R_+-R_h)^1),
\end{equation}
so that the analysis presented in the odd-parity case holds in this case too.

\section{Heun and hypergeometric equations, confluencies and superpositions}\label{appheun}

In this Appendix, we discuss the class of differential equations encoding the radial problem of the Kerr-de Sitter perturbations and of its (near-)extremal geometries. 
As first noticed in \cite{Suzuki:1998vy}, both radial and angular equations for the considered class of perturbations around Kerr-de Sitter black holes in four dimensions can be described in terms of the Heun equation.

The Heun equation is a Fuchsian differential equation having four regular singularities, located at $z=0,1,\infty,t$:
\begin{equation}\label{heuncanonical}
w''(z)+\left( \frac{\gamma}{z}+\frac{\delta}{z-1}+\frac{\epsilon}{z-t} \right)w'(z)+\frac{\alpha \beta z - q}{z(z-1)(z-t)}\,w(z) = 0.
\end{equation}
By redefining
\begin{equation}
w(z)=z^{-\gamma/2}\,(1-z)^{-\delta/2}\,(t-z)^{-\epsilon/2}\,\psi(z),
\end{equation}
the differential equation can be written in normal form as
\begin{equation}\label{heundiff}
\psi''(z) + \left[\frac{\frac{1}{4}-a_0^2}{z^2}+\frac{\frac{1}{4}-a_1^2}{(z-1)^2} + \frac{\frac{1}{4}-a_t^2}{(z-t)^2}- \frac{\frac{1}{2}-a_1^2 -a_t^2 -a_0^2 +a_\infty^2 + u}{z(z-1)}+\frac{u}{z(z-t)} \right]\psi(z)=0,
\end{equation}
where the $a_{z_i}$ parameters denote the indicial parameters of the singularities $z_i=0,1,\infty,t$, and $u$ is the accessory parameter.
The dictionary of parameters is given by
\begin{equation}\label{dictionormaltocanonical}
\begin{aligned}
&\gamma=1-2a_0,\quad\quad\delta=1-2a_1,\quad\quad\epsilon=1-2a_t,\\
&\alpha=1-a_0-a_1-a_t+a_{\infty},\quad\beta=1-a_0-a_1-a_t-a_{\infty},\\
&q=\frac{1}{2}+t\left(a_0^2+a_1^2+a_t^2-a_{\infty}^2\right)-a_t-a_1 t+a_0\left[2a_t-1+t\left(2a_1-1\right)\right]+(1-t)\,u.
\end{aligned}
\end{equation}

\begin{figure}[h!]
\centering
\includegraphics[width=10cm]{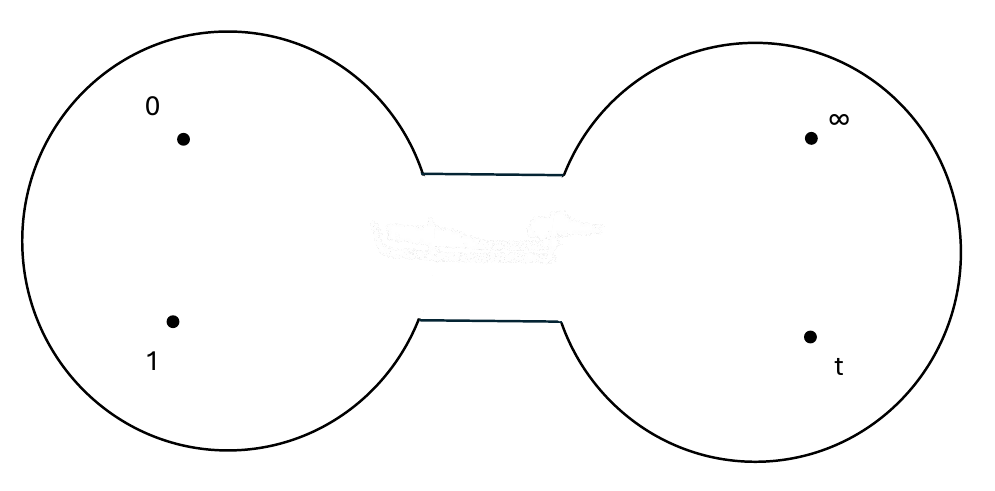}
\caption{Singularity structure of the Heun equation in the regime $|t|\gg 1$.}
\label{heunfigure}
\end{figure}

The singularity structure of the differential equation can be represented by a four-punctured sphere, whose decomposition in two joint three-punctured spheres depends on the position of the singularity at $z=t$. In this Appendix, we consider the regime in which $|t|\gg 1$, as we pictorially show in Figure \ref{heunfigure}.

The accessory parameter $u$ is related to the composite monodromy parameter $a$ via the Matone relation \cite{Matone:1995rx}
\begin{equation}\label{Matone}
u=-\frac{1}{4}+a^2-a_{\infty}^2+a_t^2+t\,\partial_tF^{N_f=4}\left(1/t\right),
\end{equation}
where $F^{N_f=4}\left(1/t\right)$ is the instanton part of the free energy of the $\mathcal{N}=2$ $SU(2)$ gauge theory with four (anti-)fundamental hypermultiplets.

The Kerr-de Sitter geometry admits different (near-)extremal regimes, whose spectral properties are encoded in differential equations that can be obtained from the Heun equation by considering confluences and/or superposition of its singularities.\footnote{In the gauge theory language, the confluence procedure corresponds to the decoupling of a hypermultiplet, whereas the superposition corresponds to the ungauging procedure.} We pictorially represented the different regimes in Figure \ref{fig:sharkfin}.

The cold geometry is obtained by taking the confluence between the inner horizon $R_i$ and the outer horizon $R_h$. The radial problem in the cold extremal regime is described by a confluent Heun equation:
\begin{equation}\label{cheuncanonical}
w''(z)+\left( \frac{\gamma}{z}+\frac{\delta}{z-1}+\epsilon \right)w'(z)+\frac{\alpha z - q}{z(z-1)}\,w(z) = 0.
\end{equation}
This has two regular singularities, at $z=0$ and $z=1$, and an irregular singularity of rank 1 at $z=\infty$.
By redefining
\begin{equation}
w(z)=e^{-\epsilon\,z/2}z^{-\gamma/2}\,(1-z)^{-\delta/2}\,\psi(z),
\end{equation}
the differential equation can be written in normal form as
\begin{equation}\label{cheundiff}
\psi''(z) + \left[\frac{\frac{1}{4}-a_0^2}{z^2}+\frac{\frac{1}{4}-a_1^2}{(z-1)^2} + \frac{u-\frac{1}{2}+a_0^2+a_1^2}{z(z-1)}+\frac{\mu\,\epsilon}{z}-\frac{\epsilon^2}{4} \right]\psi(z)=0,
\end{equation}
where 
\begin{equation}
\begin{aligned}
&\gamma=1-2a_0,\quad\quad\delta=1-2a_1,\\
&\alpha=\epsilon\left(1+\mu-a_0-a_1\right),\\
&q=(-a_0-a_1+1) (a_0+a_1)+\epsilon  \left(-a_0+\mu +\frac{1}{2}\right)-u.
\end{aligned}
\end{equation}
The equation \eqref{cheundiff} can be obtained from \eqref{heundiff} by using \eqref{Matone}, the parameterization
\begin{equation}\label{confluencelimit}
a_t=\frac{\eta -\mu }{2},\quad\quad a_{\infty}=\frac{\eta +\mu }{2},\quad\quad t=\frac{\eta }{\epsilon },
\end{equation}
and taking the limit $\eta\to\infty$. In particular, the accessory parameter $u$ appearing in \eqref{cheundiff} admits the following relation with the composite monodromy parameter of the confluent Heun equation:
\begin{equation}\label{matoneNf3}
u=\frac{1}{4}-a^2+\epsilon\,\partial_{\epsilon}F^{N_f=3}(\epsilon),
\end{equation}
where \eqref{matoneNf3} is obtained from \eqref{Matone} by using the double scaling limit \eqref{confluencelimit}.

The rotating Nariai geometry, instead, is obtained by rescaling the radial variable, zooming in the region between the approaching outer and cosmological horizons $R_h$ and $R_+$. The radial problem can be described by a hypergeometric equation
\begin{equation}\label{Hypergeometricdiffeq}
z(1-z)\, w''(z)+\left[c-(a+b+1)z\right]\,w'(z)-a\,b\, w(z) = 0.
\end{equation}
This has three regular singularities, at $z=0,1,\infty$.
By redefining 
\begin{equation}\label{hypergeq}
w(z)=z^{-\frac{c}{2}}\, (1-z)^{\frac{-a-b+c-1}{2}}\,\psi(z),
\end{equation}
it can be rewritten in normal form as
\begin{equation}\label{Hypergeometricnormal}
\psi''(z) + \left[\frac{\frac{1}{4}-a_0^2}{z^2}+\frac{\frac{1}{4}-a_1^2}{(z-1)^2} + \frac{a_0^2+a_1^2-a_{\infty}^2-\frac{1}{4}}{z(z-1)} \right]\psi(z)= 0,
\end{equation}
where
\begin{equation}
\begin{aligned}
a_0=\frac{1-c}{2},\quad a_1=\frac{c-a-b}{2},\quad a_{\infty}=\frac{b-a}{2},
\end{aligned}
\end{equation}
and inverse dictionary
\begin{equation}
\begin{aligned}
a=\frac{1}{2}-a_0-a_1-a_\infty,\quad b= \frac{1}{2}-a_0-a_1+a_\infty,\quad c= 1-2 a_0.
\end{aligned}
\end{equation}

The hypergeometric equation can be obtained from the Heun equation by considering a superposition between the singularities at $z=t$ and $z=\infty$.

In the limit $t\to\infty$, with no additional rescaling of the other parameters, the instanton corrections are suppressed, and the potential in \eqref{heundiff} reduces to
\begin{equation}
\frac{\frac{1}{4}-a_0^2}{z^2}+\frac{\frac{1}{4}-a_1^2}{(z-1)^2} + \frac{\frac{1}{4}-a_t^2}{(z-t)^2}+\frac{a_0^2+a_1^2 - a^2 -\frac{1}{4}}{z(z-1)},
\end{equation}
which coincides with \eqref{Hypergeometricnormal} with the replacement $a\to a_{\infty}$. 

The same limit (as well as the replacement $a\to a_{\infty}$) maps the connection formulae for the Heun equation between the solutions at $z=0$ and $z=1$ in the large $t$ regime \cite{Bonelli:2022ten} to the well-known connection formulae between the corresponding solutions of the hypergeometric differential equation \cite{gauss1866carl}. The same reduction applies from formula (4.1.23) in \cite{Bonelli:2022ten} to the connection formula between the hypergeometric functions centered at $z=1$ and $z=\infty$ in the limit $t\to 0$, using also the replacement $a\to a_0$.

%\begin{figure}[h!]
%\centering
%\includegraphics[width=10cm]{heunto2F12.pdf}
%\caption{Superposition of singularities from Heun to hypergeometric}
%\label{heunto2F1figure}
%\end{figure}

Finally, there is an additional (near-)extremal geometry, obtained in the regime in which all three singularities $R_i,R_h,R_+$ are close to each other. The corresponding extremal geometry is known as the ultracold geometry.
As analysed in Section \ref{sec:ultracold}, the differential equation in this near-extremal regime is obtained by rescaling twice the radial variable and reduces to a confluent hypergeometric equation
\begin{equation}\label{1F1canonical}
y\,w''(y)+(c-y)\,w'(y)-a\,w(y)=0.
\end{equation}
The corresponding equation in normal form is also known as Whittaker equation,
\begin{equation}
\psi''(y)+\left[-\frac{1}{4}+\frac{\kappa }{y}+\frac{\frac{1}{4}-\mu ^2}{y^2}\right]\psi(y)=0,
\end{equation}
which is obtained from \eqref{1F1canonical} by redefining
\begin{equation}
w(y)=e^{y/2}\, y^{-\frac{c}{2}}\,\psi(y),
\end{equation}
and where
\begin{equation}
\kappa=\frac{c-2 a}{2} \quad\quad \mu=\pm \frac{1-c}{2}.
\end{equation}
The confluent hypergeometric equation has a regular singularity at $y=0$ and an irregular singularity of rank 1 at $y=\infty$. It can be obtained from \eqref{hypergeq} by redefining the variable $z$ as $y/b$ and considering the limit $b\to\infty$.

We collected the various differential equations involved in our analysis in the confluence diagram in Figure \ref{fig:confluences}, where the cusps in the picture on the top and on the right denote the presence of an irregular singularity, whereas the circles represent the regular singularities.

\section{Kerr-anti-de Sitter in four dimensions}\label{appAdS}

The differential equation for the radial problem in the four-dimensional Kerr-anti-de Sitter (Kerr-AdS$_4$) case is the same as in the asymptotically de Sitter case, just with the sign of $\Lambda$ flipped. In this case, there is only one extremal geometry, which is the cold one analysed in Section \ref{sec:coldKerrdS} in the de Sitter case. We can still use the same variable \eqref{zKerrdScold} and the same dictionary provided in \eqref{dictioKerrdScold} for the rewriting of the differential equation, with the caveat that now $\Lambda<0$.

The difference with respect to the asymptotically de Sitter case lies in the different boundary condition satisfied by the quasinormal modes. Indeed, we impose a Dirichlet boundary condition at the AdS boundary. Since the AdS boundary is not a singularity of the differential equation for the considered class of perturbations, we first analytically continue the ingoing solution at the horizon $r=R_h$, located at $z=t$, and rewrite it as a linear combination of a basis of local solutions around a singular point such that the AdS boundary lies in the region of convergence of these local solutions. Finally, we impose the Dirichlet boundary condition. This was analysed in the non-rotating case in \cite{Aminov:2023jve}.
In particular, the quantization condition for QNMs in Kerr-AdS$_4$ reads
\begin{equation}\label{quantcondAdS}
\begin{aligned}
&\Biggl\{\left[\sum_{\sigma=\pm}\frac{\Gamma(-2\sigma a)\Gamma(2a_1)\Gamma(1-2\sigma a)t^{\sigma a}e^{-\frac{\sigma}{2}\partial_aF(t)+\frac{1}{2}\partial_{a_1}F(t)}e^{2i\pi a_1}}{\Gamma\left(\frac{1}{2}-a_t-\sigma a+a_0\right)\Gamma\left(\frac{1}{2}-a_t-\sigma a-a_0\right)\Gamma\left(\frac{1}{2}-\sigma a+a_1+a_{\infty}\right)\Gamma\left(\frac{1}{2}-\sigma a+a_1-a_{\infty}\right)}\right]\times\\
&\times\biggl(\frac{z-t}{1-t}\biggr)^{-\alpha}\mathrm{Heun}\biggl(t,q+\alpha(\delta-\beta),\alpha,\delta+\gamma-\beta,\delta,\gamma,t\frac{1-z}{t-z}\biggr)+\\
&+\left[\sum_{\sigma=\pm}\frac{\Gamma(-2\sigma a)\Gamma(-2a_1)\Gamma(1-2\sigma a)t^{\sigma a}e^{-\frac{\sigma}{2}\partial_aF(t)-\frac{1}{2}\partial_{a_1}F(t)}}{\Gamma\left(\frac{1}{2}-a_t-\sigma a+a_0\right)\Gamma\left(\frac{1}{2}-a_t-\sigma a-a_0\right)\Gamma\left(\frac{1}{2}-\sigma a-a_1+a_{\infty}\right)\Gamma\left(\frac{1}{2}-\sigma a-a_1-a_{\infty}\right)}\right]\times\\
&\times(z-1)^{1-\delta}\biggl(\frac{z-t}{1-t}\biggr)^{-\alpha-1+\delta}\\
&\times\mathrm{Heun}\biggl(t,q-(\delta-1)\gamma t-(\beta-1)(\alpha-\delta+1),-\beta+\gamma+1,\alpha-\delta+1,2-\delta,\gamma,t\frac{1-z}{t-z}\biggr)\Biggr\}\bigg|_{z=z_{\infty}}=0,
\end{aligned}
\end{equation}
where the parameters of the Heun functions can be obtained from the gravitational dictionary \eqref{dictioKerrdScold} and the dictionary \eqref{dictionormaltocanonical}.

Although the full quantization condition is much more complicated, in the near-cold regime the decoupling mechanism does not change: indeed, there will be a branch of solutions coming from the poles of the $\Gamma$-function $\Gamma\left(\frac{1}{2}-a_t-\sigma a+a_0\right)$ which is common in both connection coefficients in \eqref{quantcondAdS}.
This implies that the analysis presented in Section \ref{sec:coldKerrdS} applies to this case too, where in all expressions $\Lambda$ has to be understood as a negative parameter.

\section{Gauge theory conventions}\label{appgauge}

Let $Y$ be a Young diagram with column heights  $(Y_1\ge Y_2\ge\dots)$ and row lengths  $(Y'_1\ge Y'_2,\dots)$. For every box $s=(i,j)$ of $Y$ we define the \emph{arm length} and the \emph{leg length} as
\begin{equation}
A_Y(i, j) = Y_j -  i, \quad\quad L_Y(i, j) =Y'_i - j.
\end{equation}
We denote a pair of Young diagrams by $\vec{Y}=\left( Y_1, Y_2 \right)$ and the total number of boxes by $| \vec{Y} | = | Y_1 | + | Y_2 |$. We denote the vacuum expectation value (v.e.v.) of the scalar in the vector multiplet by $\vec{a}=(a_1,a_2)$ and the parameters characterizing the $\Omega$-background by $\epsilon_1,\epsilon_2$. We define the hypermultiplet contribution to the partition function:
\begin{equation}
\begin{aligned}
&z_{\text{hyp}} \left( \vec{a}, \vec{Y}, m \right) =\prod_{k= 1,2} \prod_{(i,j) \in Y_k} \left[ a_k + m + \epsilon_1 \left( i - \frac{1}{2} \right) + \epsilon_2 \left( j - \frac{1}{2} \right) \right] \,, 
\end{aligned}
\end{equation}
and the vector multiplet contribution:
\begin{equation}
\begin{aligned}
&z_{\text{vec}} \left( \vec{a}, \vec{Y} \right) =\prod_{i,j=1}^2\prod_{s\in Y_i}\frac{1}{a_i-a_j-\epsilon_1L_{Y_j}(s)+\epsilon_2(A_{Y_i}(s)+1)}\prod_{t\in Y_j}\frac{1}{-a_j+a_i+\epsilon_1(L_{Y_i}(t)+1)-\epsilon_2A_{Y_j}(s)}\,.
\end{aligned}
\end{equation}
We use the conventions $\epsilon_1=1$ and $\vec{a}=(a,-a)$. The instanton counting parameter is $t=e^{2\pi i\tau}$, where $\tau$ is related to the gauge coupling by $\tau=\frac{\theta}{2\pi}+i\frac{4\pi}{g_{\rm YM}^2}$.
We denote with $m_1,m_2,m_3,m_4$ the masses of the four hypermultiplets and we introduce the gauge parameters $a_0,a_t,a_1,a_{\infty}$ satisfying, in our notation and with the assumption $|t|\ll 1$,
\begin{equation}\label{gaugemasses}
\begin{aligned}
m_1&=a_t-a_0,\quad\quad\  
m_2=a_t+a_0,\\
m_3&=a_{\infty}+a_1,\quad\quad
m_4=-a_{\infty}+a_1.
\end{aligned}
\end{equation}
In this small $t$ regime, the instanton part of the Nekrasov-Shatashvili free energy is given as a power series in $t$ by
\begin{equation}
\begin{aligned}
F(t)=&\lim_{\epsilon_2\to 0}\epsilon_2\log\Biggl[\left(1-t\right)^{-2\epsilon_2^{-1}\left(\frac{1}{2}+a_1\right)\left(\frac{1}{2}+a_t\right)}\,\sum_{\vec{Y}}t^{|\vec{Y}|}z_{\text{vec}} \left( \vec{a}, \vec{Y} \right)\prod_{i=1}^4z_{\text{hyp}} \left( \vec{a}, \vec{Y}, m_i \right)\Biggr].
\end{aligned}
\end{equation}
The parameter $a$ characterizes the composite monodromy around $z=0$ and $z=t$ in the 4-punctured sphere. From the gauge theory point of view, $a$ is expressed in a series expansion in $t$, obtained by inverting the Matone relation \cite{Matone:1995rx}
\begin{equation}
u =-\frac{1}{4} - a^2 + a_t^2 + a_0^2 + t\, \partial_t F(t),
\end{equation}
where the parameter $u$ is the complex modulus parametrizing the energy of the corresponding Seiberg-Witten curve \cite{seiberg1994a,seiberg1994}, and corresponds to the accessory parameter of the Heun equation.
Explicitly, the expansion up to the first instanton contribution of $a$ reads
\begin{equation}\label{agaugeinstanton}
\begin{aligned}
a=\pm\Biggl\{\sqrt{-\frac{1}{4}-u+a_t^2+a_0^2}+\frac{\bigl(\frac{1}{2}+u-a_t^2-a_0^2-a_1^2+a_{\infty}^2\Bigr)\Bigl(\frac{1}{2}+u-2a_t^2\Bigr)}{2(1+2u-2a_t^2-2a_0^2)\sqrt{-\frac{1}{4}-u+a_t^2+a_0^2}}t+\mathcal{O}(t^2)\Biggr\}.
\end{aligned}
\end{equation}

A similar construction holds for the $\mathcal{N}=2$ quiver gauge theory with group $SU(2)\times SU(2)$ and with four (anti-)fundamental hypermultiplets. This is the gauge theory relevant for the differential equation for even-parity gravitational perturbations around Schwarzschild-de Sitter black holes analysed in Appendix \ref{appzerilli}. We refer to \cite{Arnaudo:2025kof} for the notations and conventions used. 

 \bibliographystyle{JHEP}
 \bibliography{biblio}

\end{document}